\pdfoutput=1
\documentclass[11pt,twoside,a4paper,cmspaper,final,collab]{cms-tdr}
\def\svnVersion{edbd196-D}\def\svnDate{2021/05/30}\def\cmsCernNoTag{CERN-EP-2021-028}\def\cmsCernDate{\today}\def\cmsMessage{"Published in the Journal of High Energy Physics as \href{http://dx.doi.org/10.1007/JHEP05(2021)182}{\doi{10.1007/JHEP05(2021)182}.}"}
\begin{document}\cmsNoteHeader{HIN-18-003}

\newcommand{\mupmum}{\ensuremath{\PGmp\PGmm}\xspace}
\newcommand{\emu}{\ensuremath{\Pe\PGm}\xspace}
\newcommand{\mmumu}{\ensuremath{m_{\PGm\PGm}}\xspace}
\newcommand{\DY}{\ensuremath{\PZ/\PGg^{*}}\xspace} 
\newcommand{\etalab}{\ensuremath{\eta_{\text{lab}}}\xspace}
\newcommand{\ycm}{\ensuremath{y_{\mathrm{CM}}}\xspace}
\newcommand{\phistar}{\ensuremath{\phi^*}\xspace}
\newcommand{\rfb}{\ensuremath{R_{\mathrm{FB}}}\xspace}
\newcommand{\etacm}{\ensuremath{\eta_{\mathrm{CM}}}\xspace}
\newcommand{\fig}[1]{Fig.~\ref{#1}\xspace}
\newcommand{\tab}[1]{Table~\ref{#1}\xspace}
\newcommand{\myPb}{\ensuremath{\mathrm{Pb}}\xspace}
\newcommand{\pp}{{\ensuremath{\Pp\Pp}}\xspace}
\newcommand{\pPb}{\ensuremath{\Pp\myPb}\xspace}
\newcommand{\EPOS} {{\textsc{epos}}\xspace}
\newcommand{\PYQUEN} {{\textsc{pyquen}}\xspace}
\newcommand{\Irel}{\ensuremath{I_{\text{rel}}}\xspace}
\newcommand{\mylumi}{\ensuremath{173.4 \pm 6.1\nbinv}\xspace}
\newcommand{\sqrts}{\ensuremath{\sqrt{s}}\xspace}

\cmsNoteHeader{HIN-18-003} 
\title{Study of Drell--Yan dimuon production in proton-lead collisions at \texorpdfstring{$\sqrtsNN = 8.16\TeV$}{sqrt(s_NN) = 8.16 TeV}}

\date{\today}

\abstract{
Differential cross sections for the Drell--Yan process, including \PZ boson production, using the dimuon decay channel are measured in proton-lead (\pPb) collisions at a nucleon-nucleon centre-of-mass energy of 8.16\TeV. A data sample recorded with the CMS detector at the LHC is used, corresponding to an integrated luminosity of 173\nbinv.     The differential cross section as a function of the dimuon mass is measured in the range 15--600\GeV, for the first time in proton-nucleus collisions.    It is also reported as a function of dimuon rapidity over the mass ranges 15--60\GeV and 60--120\GeV,    and ratios for the \Pp-going over the \myPb-going beam directions are built.    In both mass ranges, the differential cross sections as functions of the dimuon transverse momentum \pt and of a geometric variable \phistar are measured, where \phistar highly correlates with \pt but is determined with higher precision.    In the \PZ mass region, the rapidity dependence of the data indicate a modification of the distribution of partons within a lead nucleus as compared to the proton case. The data are more precise than predictions based upon current models of parton distributions. 
}

\hypersetup{%
pdfauthor={CMS Collaboration},%
pdftitle={Study of Drell-Yan dimuon production in proton-lead collisions at sqrt(s[NN]) = 8.16 TeV},%
pdfsubject={CMS},%
pdfkeywords={CMS,  relativistic heavy ion physics, Drell-Yan, nuclear parton distribution functions, pPb, 8.16 TeV}}

\maketitle 

\section{Introduction}

The annihilation of a quark-antiquark pair into two oppositely charged leptons, through the exchange of a \PZ boson or a virtual photon (\DY) in the $s$-channel, 
is known as the Drell--Yan (DY) process~\cite{Drell:1970wh}. The theoretical derivation of the matrix elements is 
available up to next-to-next-to-leading order in perturbative quantum
chromodynamics (QCD) with next-to-leading order (NLO) electroweak (EW) corrections~\cite{QCDNNLO, DYNNLO, DYNNLO1, DY-Theory}. A precise measurement of this process can add valuable information on its
nonperturbative part, including the effect of parton distribution functions (PDFs)~\cite{Martin:1994kn}.

Measurements of EW bosons in proton-nucleus and nucleus-nucleus collisions probe the nuclear modification of the PDFs~\cite{Kartvelishvili:1995fr,Vogt:2000hp,Zhang:2002yz,Paukkunen:2010qg}. The presence of a nuclear environment has been long observed~\cite{Aubert:1983xm} to modify the parton densities in the nucleus, as compared to those in a free nucleon. A first-principle description of such (nonperturbative) nuclear effects remains an open challenge, but they can be modelled using nuclear PDFs (nPDFs) determined with data in the same collinear factorisation approach as for free protons. Global fits of nPDFs~\cite{deFlorian:2011fp,Khanpour:2016pph,epps16,Kovarik:2015cma,Walt:2019slu,AbdulKhalek:2019mzd,AbdulKhalek:2020yuc,Kusina:2020lyz}
predict a suppression for small longitudinal momentum fraction $x$, $x \lesssim 10^{-2}$ (\ie shadowing~\cite{Armesto:2006ph} region), and an enhancement for intermediate $x$, $10^{-2} \lesssim x \lesssim 10^{-1}$ (\ie antishadowing region).

Many measurements of the DY process, including the mass dependence, have been performed in proton-proton (\pp) collisions, for instance by the ATLAS~\cite{Aad:2013iua,Aad:2014qja,Aad:2016zzw,Aad:2019bdc,Aad:2019wmn}, CMS~\cite{Chatrchyan:2011cm,Chatrchyan:2013tia,CMS:2014jea,Sirunyan:2018owv,Sirunyan:2019bzr}, and PHENIX~\cite{Aidala:2018ajl} experiments.
Measurements of the \PZ boson production have been performed in proton-lead (\pPb) collisions by the ALICE~\cite{Alice:2016wka,Acharya:2020puh}, ATLAS~\cite{Aad:2015gta}, and CMS~\cite{hin-15-002} experiments, as functions of rapidity, transverse momentum, or centrality (related to the impact parameter of the collision).

In this paper, we report the measurement of the differential cross section for \mupmum production via the DY process, as a function of the following variables:

\begin{itemize}
 \item dimuon mass, \mmumu, in the interval $15<\mmumu<600\GeV$;
 \item dimuon transverse momentum, \pt, in two dimuon mass intervals (15--60\GeV and 60--120\GeV, targeting the continuum at low mass and the \PZ boson, respectively);
 \item dimuon rapidity in the nucleon-nucleon centre-of-mass (CM) frame, \ycm, in the same two mass intervals; and
 \item \phistar~\cite{Banfi:2010cf,Banfi:2012du,Marzani:2013nza} (defined below) in the same two mass intervals.
\end{itemize}

The dimuon mass and \phistar dependencies as well as cross sections in the dimuon mass range 15--60\GeV are reported for the first time in proton-nucleus collisions.

The variable \phistar, used in numerous \PZ boson studies, is defined as
\begin{linenomath}
\begin{equation} 
 \phistar \equiv \tan \left( \frac{\pi - \Delta\phi}{2} \right) \sin (\theta^*_\eta),
 \label{eq:phistar}
\end{equation}
\end{linenomath}
where $\Delta\phi$ is the opening angle between the leptons, defined as the difference of their azimuthal angles in the plane transverse to the beam axis, and $\theta^*_\eta$ is related to the emission angle of the dilepton system with respect to the beam. The variable $\theta^*_\eta$ is defined in a frame that is Lorentz-boosted along the beam direction such that the two leptons are back-to-back in the transverse plane. This angle $\theta^*_\eta$ is related to the pseudorapidities of the leptons by the relation
\begin{linenomath}
\begin{equation}
 \cos (\theta^*_\eta) = \tanh ( \Delta\eta /2 ),
\end{equation}
\end{linenomath}
where $\Delta\eta$ is the difference in pseudorapidity between the two leptons. By construction, \phistar is greater than zero. This quantity strongly correlates with the dimuon \pt, while only depending on angular quantities for the leptons. Thus, it is measured with better precision than \pt, especially at low \pt values.
Since $\phistar \sim \pt / m$, where $m$ is the mass of the dilepton system, the range $\phistar < 1$ corresponds to dilepton \pt
up to about 100\GeV for a dilepton mass close to that of the \PZ boson.

The outline of this paper is as follows. In Section~\ref{sec:expmeth}, the experimental methods are described, from the data and simulation samples used, up to the data analysis description and systematic uncertainties estimation. Results are presented and discussed in Section~\ref{sec:results}, before the summary in Section~\ref{sec:summary}.

\section{Experimental methods} \label{sec:expmeth}

\subsection{Data taking conditions and the CMS detector}

The results reported in this paper use \pPb collision data taken by CMS at the end of 2016, at a nucleon-nucleon CM energy of $\sqrtsNN = 8.16\TeV$ at the CERN LHC. The total integrated luminosity corresponds to \mylumi~\cite{LUM17002}.
In the first part of the \pPb run, corresponding to $63 \pm 2$\nbinv, the proton beam was heading toward negative $\eta$, according to the CMS detector convention~\cite{Chatrchyan:2008zzk}, with an energy of 6.5\TeV, and colliding with a lead nucleus beam with an energy of 2.56\TeV per nucleon. The beams were swapped for the second part of the run, corresponding to $111 \pm 4$\nbinv.
Because of the asymmetric collision system, massless particles produced in the nucleon-nucleon CM frame at a given \etacm are reconstructed at $\etalab = \etacm - 0.465$ in the laboratory frame used in this paper, in which the proton is heading toward positive $\eta$. The measurements presented here are expressed in terms of \ycm.

The central feature of the CMS apparatus is a superconducting solenoid of 6\unit{m} internal diameter, providing a magnetic field of 3.8\unit{T}. Within the solenoid volume are a silicon pixel and strip tracker, a lead tungstate crystal electromagnetic calorimeter (ECAL), and a brass and scintillator hadron calorimeter (HCAL), each composed of a barrel and two endcap sections. Forward calorimeters extend the $\eta$ coverage provided by the barrel and endcap detectors. 
The hadron forward (HF) calorimeter uses steel as the absorber and quartz fibres as the sensitive material. The two halves of the HF are located 11.2\unit{m} from the interaction region, one on each end, and together they provide coverage in the range $3.0 < \abs{\eta} < 5.2$. They also serve as luminosity monitors. 
Muons are measured in the range $\abs{\eta} < 2.4$ in gas-ionisation chambers embedded in the steel flux-return yoke outside the solenoid, with detection planes made using three technologies: drift tubes, cathode strip chambers, and resistive-plate chambers. 

Events of interest are selected using a two-tiered trigger system. The first level, composed of custom hardware processors, uses information from the calorimeters and muon detectors to select events at a rate of around 100\unit{kHz} within a fixed latency of about 4\mus~\cite{Sirunyan:2020zal}. The second level, known as the high-level trigger (HLT), consists of a farm of processors running a version of the full event reconstruction software optimised for fast processing, and reduces the event rate to around 1\unit{kHz} (up to around 20\unit{kHz} during the \pPb data taking) before data storage~\cite{Khachatryan:2016bia}. 

The reconstructed vertex with the largest value of summed physics-object $\pt^2$ is taken to be the primary \pPb interaction vertex. The physics objects are the jets, clustered using the jet finding algorithm~\cite{Cacciari:2008gp,Cacciari:2011ma} with the tracks assigned to the vertex as inputs, and the associated missing transverse momentum, taken as the negative vector sum of the \pt of those jets. During the data taking, the average number of collisions per bunch crossing was 0.18. The stability of the results has been checked against different such average number conditions.

The particle-flow algorithm~\cite{CMS-PRF-14-001} aims to reconstruct and identify each individual particle in an event, with an optimised combination of information from the various elements of the CMS detector. The energy of photons is obtained from the ECAL measurement. The energy of electrons is determined from a combination of the electron momentum at the primary interaction vertex as determined by the tracker, the energy of the corresponding ECAL cluster, and the energy sum of all bremsstrahlung photons spatially compatible with originating from the electron track. The energy of muons is obtained from the curvature of the corresponding track. The energy of charged hadrons is determined from a combination of their momentum measured in the tracker and the matching ECAL and HCAL energy deposits, corrected for zero-suppression effects and for the response function of the calorimeters to hadronic showers. Finally, the energy of neutral hadrons is obtained from the corresponding corrected ECAL and HCAL energies.

Matching muons to tracks measured in the silicon tracker results in a relative transverse momentum resolution, for muons with \pt up to 100\GeV, of 1\% in the barrel and 3\% in the endcaps. The \pt resolution in the barrel is better than 7\% for muons with \pt up to 1\TeV~\cite{Sirunyan:2018fpa}. 

A more detailed description of the CMS detector, together with a definition of the coordinate system used and the relevant kinematic variables, can be found in Ref.~\cite{Chatrchyan:2008zzk}. 

\subsection{Simulated samples}
\label{sec:samples}

The signal and most backgrounds are modelled using Monte Carlo (MC) simulated samples. The following processes are considered: DY to \mupmum (signal) and to $\PGtp\PGtm$ (treated as background), \ttbar, diboson ($\PW\PW$, $\PW\PZ$, and $\PZ\PZ$), and single top quark production ($\PQt\PW$ and $\PAQt\PW$, collectively referred to as $\PQt\PW$ in the paper). Additional MC samples are used, for the production of \PW bosons (decaying to muon and neutrino, or \PGt lepton and neutrino) and QCD multijet events. These backgrounds are estimated using control samples in data, as described later in the text, and the MC samples are only used for complementary studies.

The DY, \PW boson, \ttbar, and $\PQt\PW$ MC samples are generated using the NLO generator \POWHEG v2~\cite{Nason:2004rx,Frixione:2007vw,Alioli:2010xd,Alioli:2008gx}, modified to account for 
the mixture of proton-proton and proton-neutron interactions occurring in \pPb collisions. The CT14~\cite{ct14} PDF set is used, with nuclear modifications from EPPS16~\cite{epps16}
for the lead nucleus. Parton showering is performed by \PYTHIA~8.212~\cite{pythia82} 
with the CUETP8M1 underlying event (UE) tune~\cite{Khachatryan:2015pea}. The decay of \PGt leptons in the $\PW\to\PGt\PGnGt$ MC samples is handled in \POWHEG using \TAUOLA 1.1.5~\cite{Jadach:1990mz}, 
including final-state radiative (FSR) quantum electrodynamics corrections using \PHOTOS 2.15~\cite{Golonka:2005pn}. The diboson and QCD multijet samples are generated at leading order using \PYTHIA.

The aforementioned event generators only simulate single proton-nucleon interactions, with the proportion of protons and neutrons found in \myPb nuclei. To consider a more realistic distribution of the UE present in \pPb collisions, simulated events are embedded into two separate
samples of minimum bias (MB) events generated with \EPOS LHC (v3400)~\cite{Pierog:2013ria}, one for each \pPb boost direction. The \EPOS MC samples provide a good description of the global event 
properties of the MB \pPb data, such as the $\eta$ distributions of charged hadrons~\cite{Sirunyan:2017vpr} and the transverse energy density~\cite{Sirunyan:2018nqr}.

A difference is found between the dimuon \pt in \POWHEG MC and that observed in data. To improve the modelling in the simulation, 
the \POWHEG \DY samples are reweighted event-by-event using an empirical function of the generated boson \pt. 
This weight is applied in \DY MC samples in the derivation of the various corrections described below. However, it is not applied in the figures of this paper, where the original \pt spectrum from \POWHEG is used.

The full detector response is simulated for all MC samples, using \GEANTfour~\cite{Agostinelli:2002hh}, with alignment and calibration conditions tuned to match collision data,  and a realistic description of the beam spot. The trigger decisions are also emulated, and the MC events 
are reconstructed with the standard CMS \pp reconstruction algorithms used for the 2016 data.

The \DY, \PW, and $\PQt\PW$ samples are normalised to their NLO cross sections provided by \POWHEG for \pPb collisions, including EPPS16 modifications. The diboson samples are normalised to the cross sections measured by the CMS Collaboration 
in \pp collisions at $\sqrts = 8\TeV$~\cite{cms_ppWW8tev,cms_ppWZ8tev,cms_ppZZ8tev}. The small difference in CM energy with the \pPb data is covered by the data-driven correction described in Section~\ref{sec:bkg} and smaller than the associated systematic uncertainty. The \ttbar background is normalised to the CMS measurement in \pPb collisions at
$\sqrtsNN = 8.16\TeV$~\cite{Sirunyan:2017xku}. All backgrounds receive a correction based on control samples in data, as described in Section~\ref{sec:bkg}.

Simulated events do not feature the same event activity (charged-particle multiplicity or energy density) as the data, mostly because selecting two energetic muons favours higher-activity events (with a larger number of binary nucleon-nucleon collisions), while the \EPOS sample used
for embedding simulates MB events. To ensure a proper description of event activity in simulation, the distribution of the energy deposited in both sides of the HF calorimeter
is reweighted event-by-event so that it matches that observed in data (selecting $\PZ\to\mupmum$ events). The corresponding weights have a standard deviation of 0.27 for a mean of 1.

\subsection{Object reconstruction and event selection}

The events used in the analysis are selected with a single-muon trigger, requiring $\pt>12\GeV$ for the muon reconstructed by the HLT. 
During both online and offline muon reconstruction, the data from the muon detectors are matched and
fitted  to  data  from  the  silicon  tracker  to  form  muon  candidates.   
Each  muon  is  required  to  be  within  the  geometrical  acceptance  of the detector, $\abs{\etalab}<2.4$.
The leading muon (with highest \pt) is matched to the HLT trigger object and
is required to have $\pt>15\GeV$, in the plateau of the trigger efficiency (around 95\%, depending on \etalab). A looser selection of $\pt>10\GeV$ is applied to the other muon.

Muons are selected by applying the standard ``tight'' selection criteria~\cite{Sirunyan:2018fpa} used, \eg in Refs.~\cite{Sirunyan:2019dox,Sirunyan:2017xku}, with an efficiency of about 98\%.
Requirements on the impact parameter and the opening angle between the two muons
are further imposed to reject cosmic ray muons.
Events are selected for further analysis if they
contain pairs of oppositely charged muons meeting the above requirements.
The $\chi^2$ divided by the number of degrees of freedom (dof) from a fit to the dimuon vertex must be smaller than 20, ensuring that the two muon tracks originate from a common vertex, thus reducing the contribution from heavy-flavour meson decays.   
In the rare events (about 0.4\%) where more than one selected dimuon pair is found, the candidate with the smallest dimuon vertex $\chi^2$ is kept.

To further suppress 
the background contributions due to muons originating from light and heavy flavour hadron decays,
muons are required to be isolated, based on the \pt sum of the charged-particle tracks around the muon.  Isolation sums are evaluated in a circular region
of the $(\eta, \phi)$ plane around the lepton candidate with $\Delta R < 0.3$, where $\Delta R = \sqrt{\smash[b]{(\Delta\eta)^2+(\Delta\phi)^2}}$. 
The relative isolation \Irel, obtained by dividing this isolation sum by the muon \pt, is required to be below 0.2.

In addition to the DY process, lepton pairs can also be produced through photon interactions. Exclusive coherent photon-induced dilepton production is enhanced in \pPb collisions compared to \pp data, because of the large charge of the lead nucleus. 
Hadronic collisions are selected by requiring at least one HF calorimeter tower with more than 3\GeV of total energy on either side of the interaction point. In order to further suppress the photon-induced background, characterised by almost back-to-back muons, events are required to contain at least one additional reconstructed track, which completely removes this background. Incoherent photon-induced dimuon production, where the photon is emitted from a parton instead of the whole nucleus and amounting to less than 5\% of the total dilepton cross section according to studies in \pp collisions at $\sqrts = 13\TeV$~\cite{Sirunyan:2018owv,Bourilkov:2016oet}, is considered part of the signal and is neither removed nor subtracted.

\subsection{Background estimation}
\label{sec:bkg}

Various backgrounds are estimated using one of the techniques described below, depending on the nature of the respective background process. Processes involving two isolated muons, such as $\DY \to \PGtp\PGtm$, \ttbar, $\PQt\PW$, and dibosons, are estimated from 
simulation and corrected using the ``\emu method''. Processes with one or more muons in jets, namely {\PW}+jets and multijet, are estimated using the ``misidentification rate method''. 

The \emu method takes advantage of the fact that the EW backgrounds, as opposed to the $\DY\to\mupmum$ signal, also contribute to the \emu final state. Events with exactly one
electron and one muon of opposite charge are used, where the muon is selected as described previously, matched to the HLT trigger muon and with $\pt>15\GeV$, while the electron~\cite{CMS:EGM-14-001} must have $\pt>20\GeV$ and fulfil the same isolation requirement as the muon. The small contribution from heavy-flavour meson decays is estimated from same-sign \emu events. 
The data-to-simulation ratio with this selection, in each bin of the measured variables, is used to correct the simulated samples in the \mupmum final state. This ratio is compatible with unity in most bins.

The misidentification rate method estimates the probability for a muon inside a jet and passing the tight selection criteria to pass the isolation requirements. This probability (the misidentification rate) is estimated as a function of \pt, separately for $\abs{\etalab}<1.2$  and $\abs{\etalab}>1.2$. A sideband in data is selected from opposite-sign dimuon events in which the dimuon vertex $\chi^2$ selection has been inverted. This sample is dominated by contributions from multijet and {\PW}+jets production, and the small contribution from EW processes, estimated using simulation, is removed.
The misidentification rate is then applied to a control dimuon data sample, passing the dimuon vertex $\chi^2$ selection but in which neither of the two muons passes the isolation requirement, to obtain the multijet contribution in the signal region, where both muons are isolated. The {\PW}+jets contribution is estimated with a similar procedure, using events in which exactly one of the two muons passes the isolation requirement. The small contribution from EW processes to these control data samples is estimated using simulation and removed. The multijet contribution in the sample with exactly one isolated muon is also accounted for, using the same technique. The validity of this method is checked in a control sample of same-sign dimuon data, which is also dominated by the multijet and {\PW}+jets processes. The same-sign data are found to be compatible with the predictions from the misidentification rate method in most bins, and the residual difference is accounted for as a systematic uncertainty.

\begin{figure}[htbp]
 \centering
\includegraphics[width=0.7\textwidth]{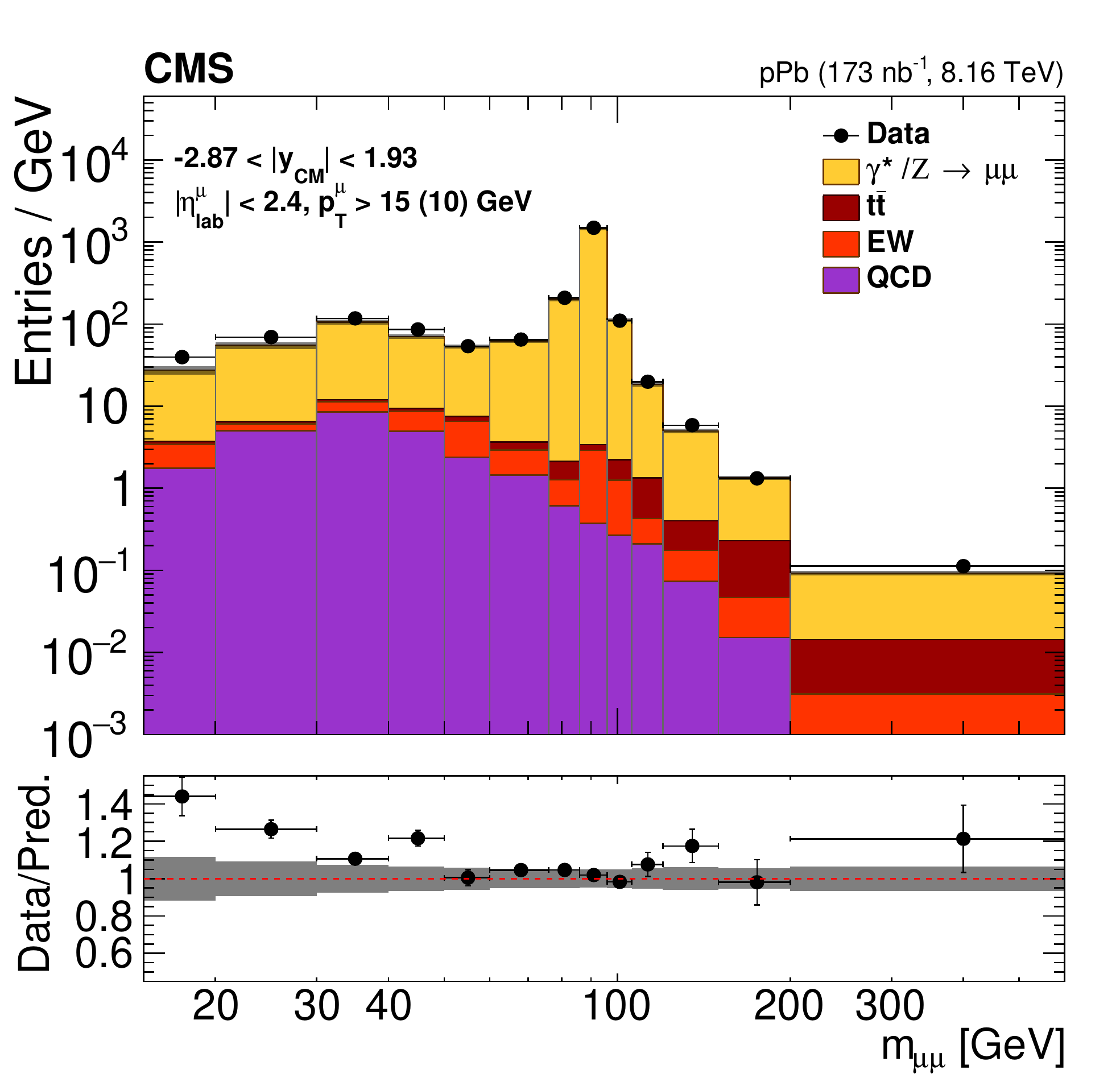}

\includegraphics[width=0.49\textwidth]{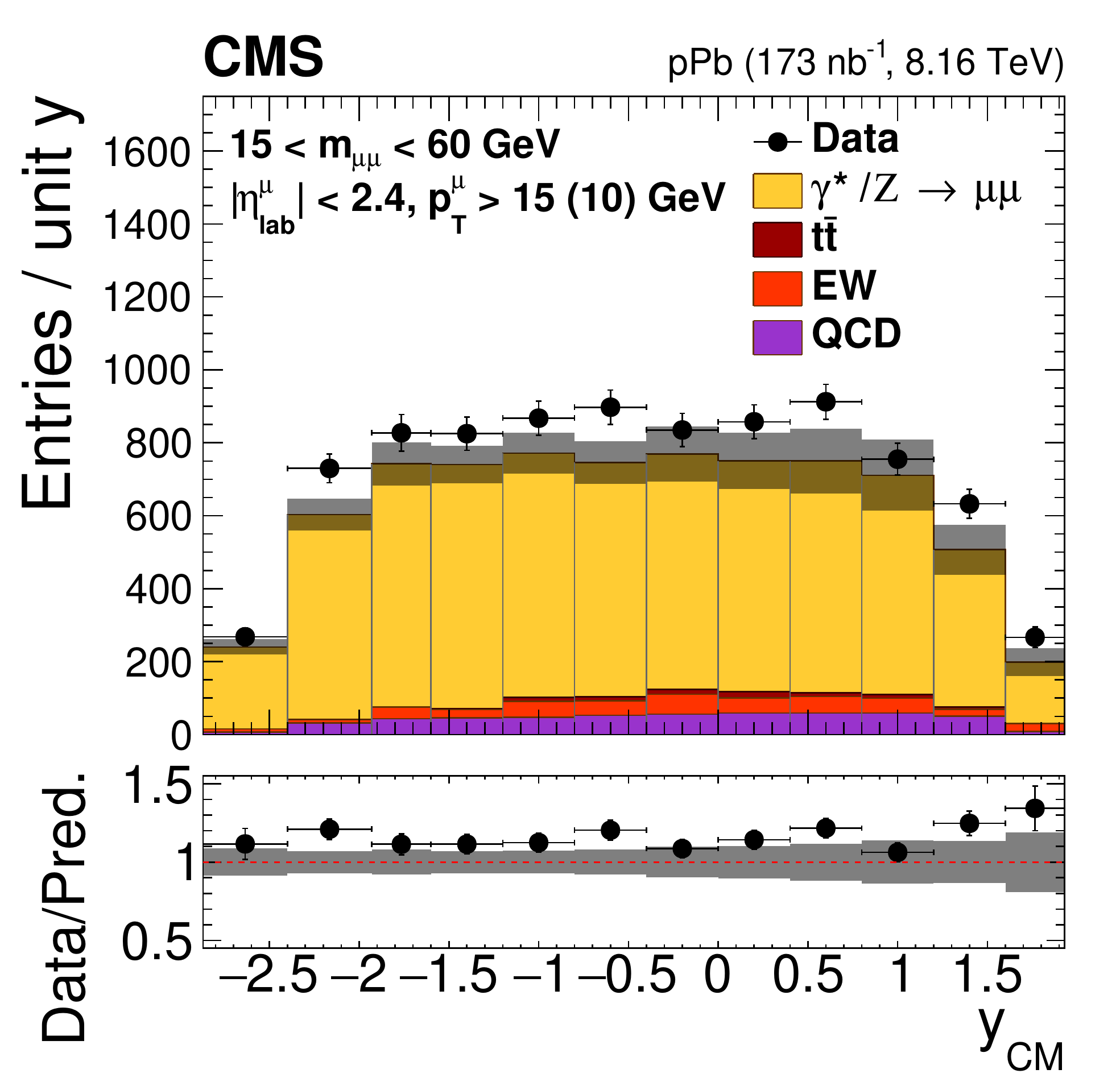} 
\includegraphics[width=0.49\textwidth]{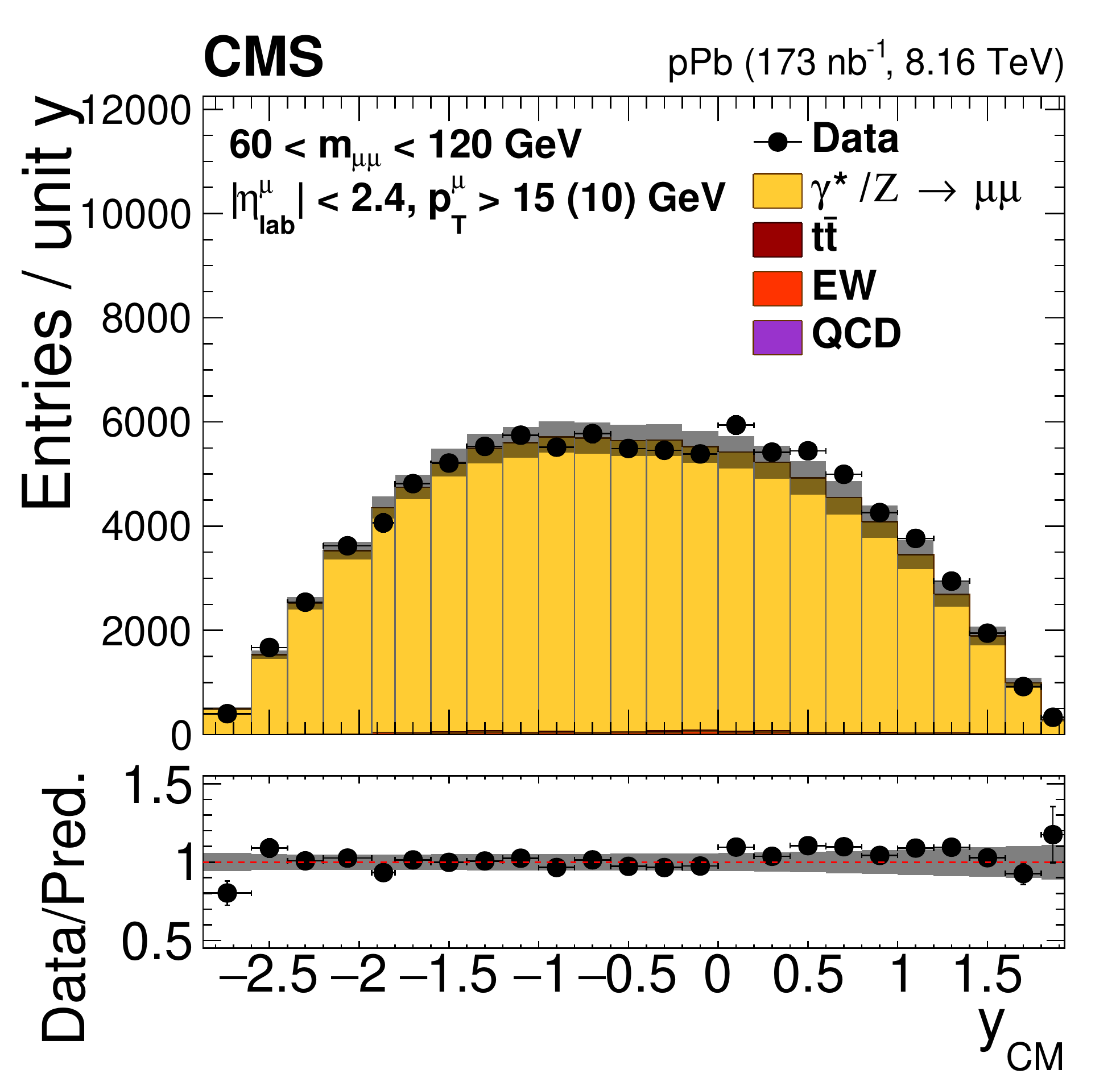}
 
\caption{
\label{fig:dataMC_mass}
Comparison of the data (black points) with the \DY signal and background expectations (filled histograms, where "EW" includes $\DY \to \PGtp\PGtm$ and diboson), estimated as described in the text,
   as a function of invariant mass (upper) and rapidity in the centre-of-mass frame for $15<\mmumu<60\GeV$ (lower left) and $60<\mmumu<120\GeV$ (lower right). Vertical error bars represent statistical uncertainties. The ratios of data over expectations are shown in the lower panels. The boson \pt reweighting described in the text is not applied. The shaded regions show the quadratic sum of the systematic uncertainties (including the integrated luminosity, but excluding acceptance and unfolding uncertainties) and the nPDF uncertainties (CT14+EPPS16).
}
\end{figure}

\begin{figure}[htbp]

\includegraphics[width=0.49\textwidth]{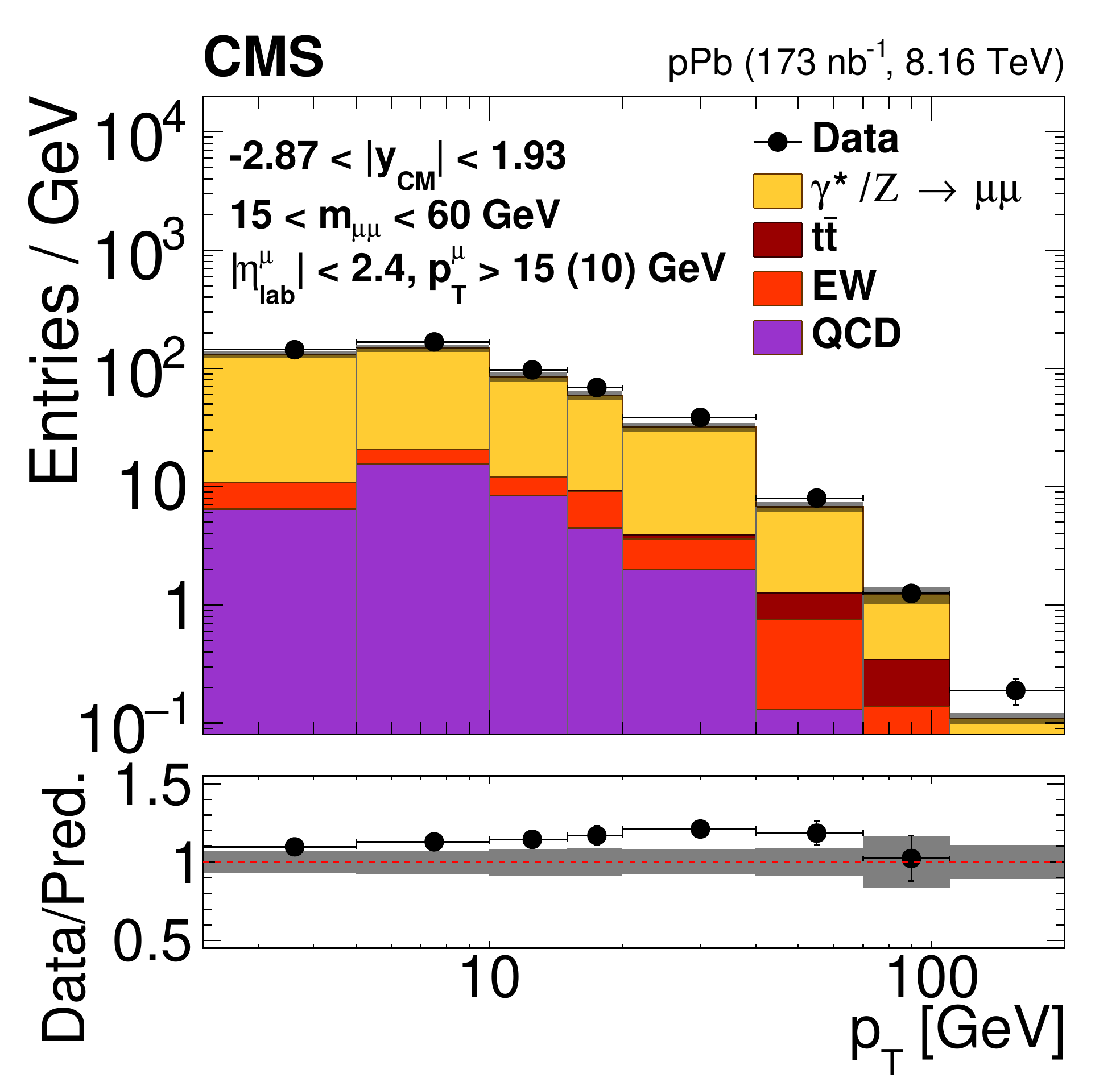} 
\includegraphics[width=0.49\textwidth]{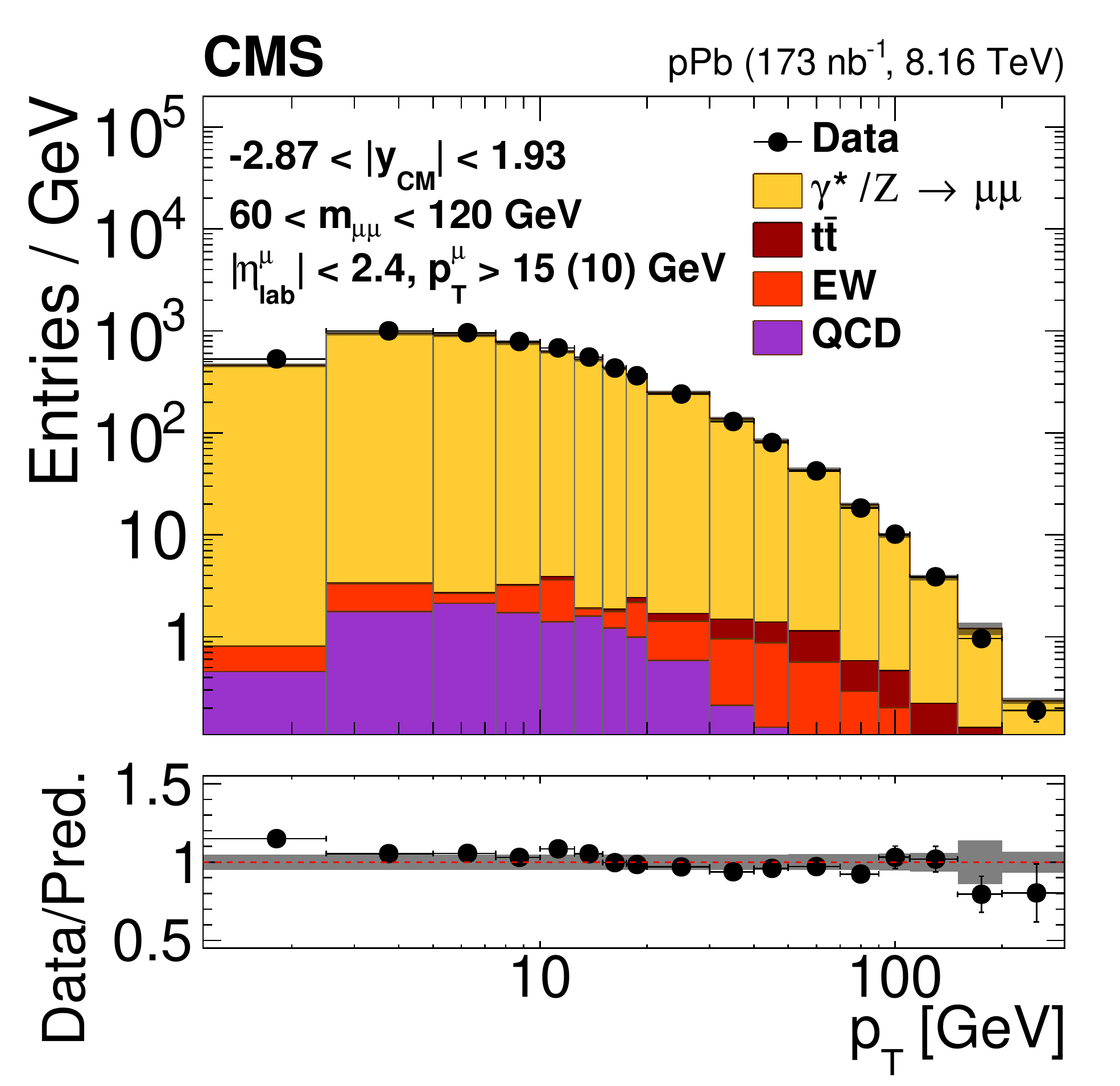}

\includegraphics[width=0.49\textwidth]{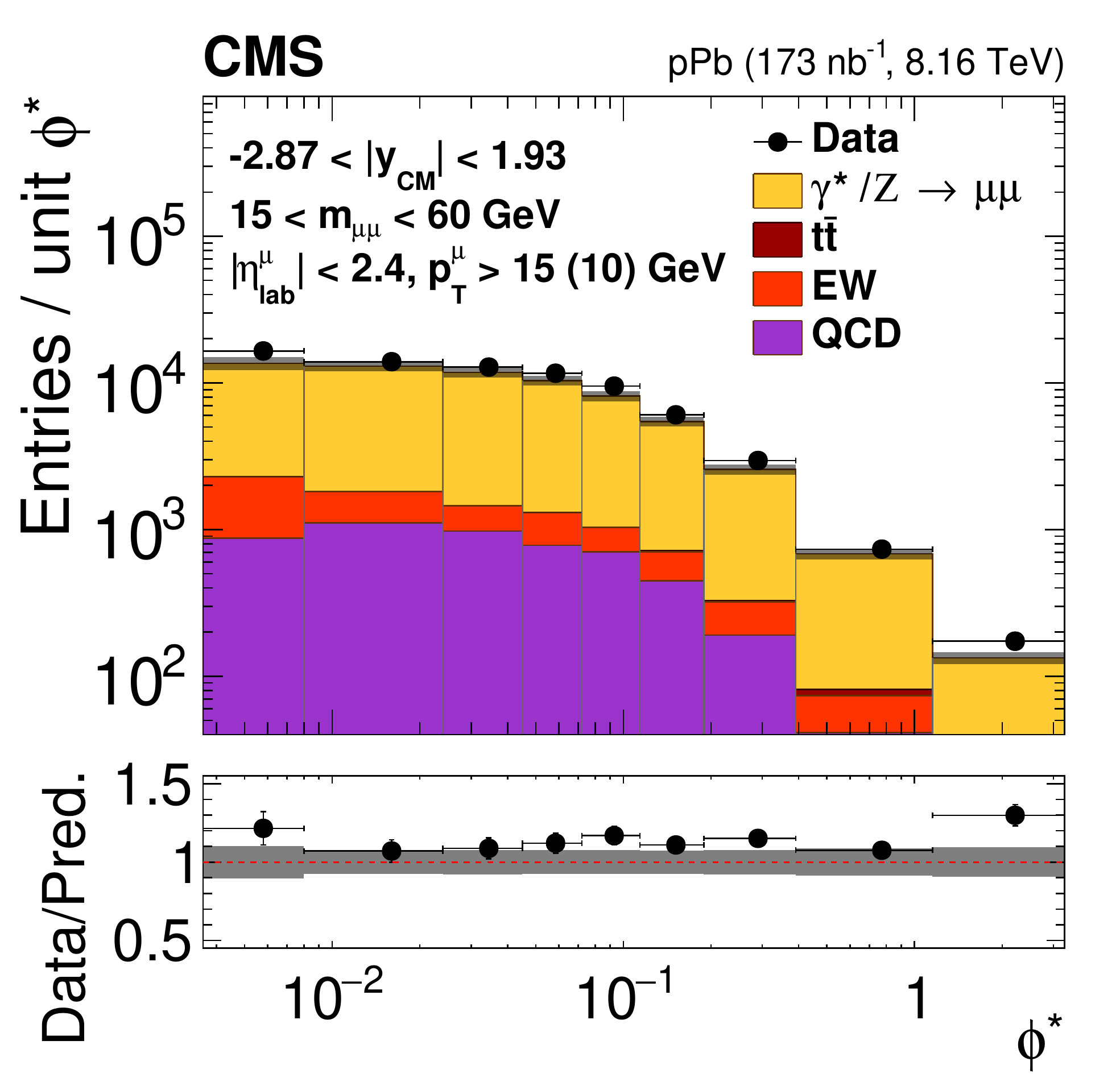} 
\includegraphics[width=0.49\textwidth]{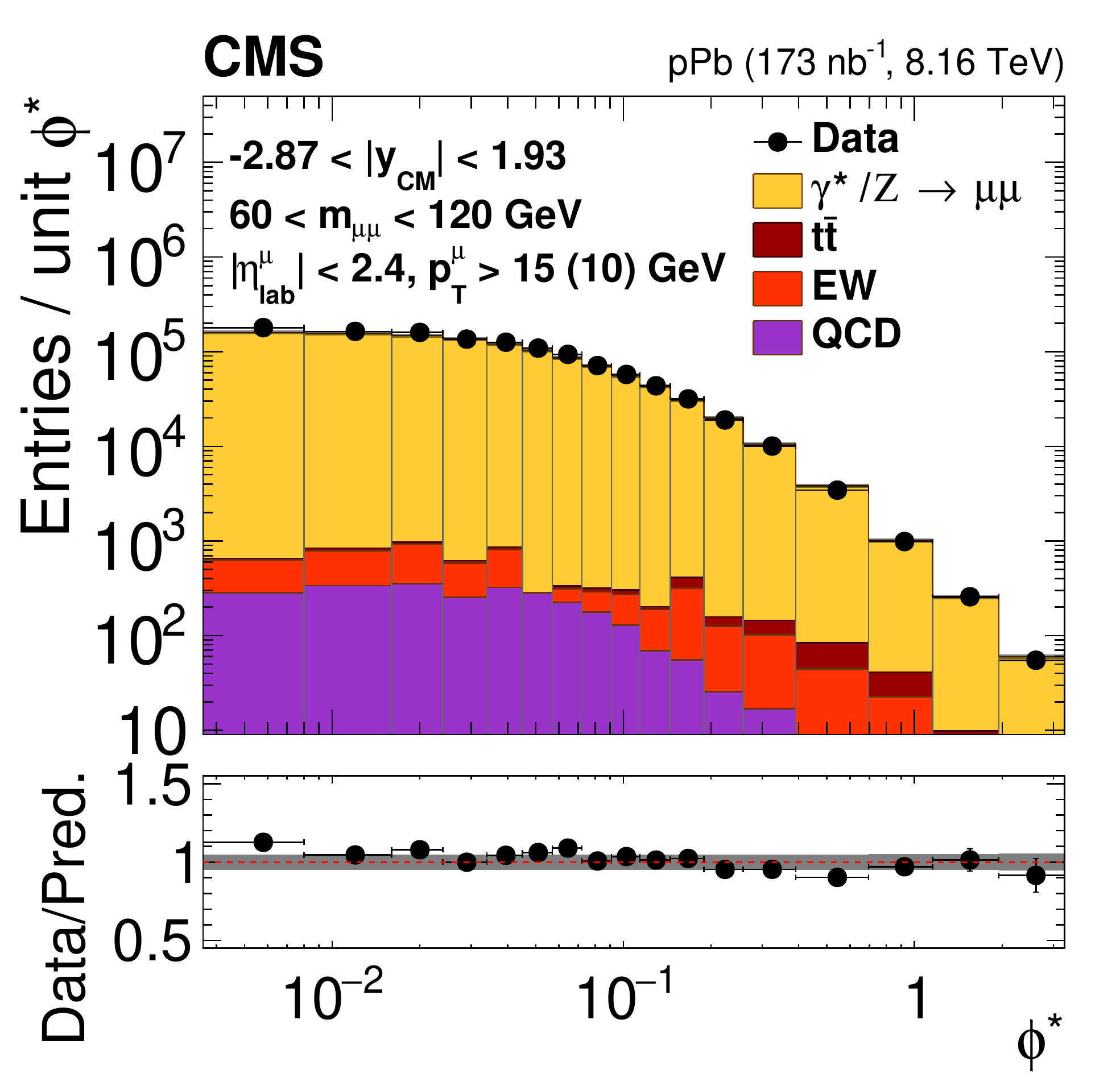} 
\caption{
\label{fig:dataMC_other}
Comparison of the data (black points) with the \DY signal and background expectations (filled histograms, where "EW" includes $\DY \to \PGtp\PGtm$ and diboson), estimated as described in the text,
as a function of \pt (upper row) and \phistar (lower row), for $15<\mmumu<60\GeV$ (left) and $60<\mmumu<120\GeV$ (right). The first bins of the \pt and \phistar distributions start at 0. Vertical error bars represent statistical uncertainties. The ratios of data over expectations are shown in the lower panels. The boson \pt reweighting described in the text is not applied. The shaded regions show the quadratic sum of the systematic uncertainties (including the integrated luminosity, but excluding acceptance and unfolding uncertainties) and the nPDF uncertainties (CT14+EPPS16).
}
\end{figure}

In Figs.~\ref{fig:dataMC_mass} and \ref{fig:dataMC_other}, data are compared to the prediction from DY simulation and background expectations estimated using the techniques described above. A good
overall agreement is found between the data and the expectation, which is dominated by the DY signal. Some hints for the differences will be discussed in terms of potential physics implications in Section~\ref{sec:results}: they include data above expectation for $\mmumu<50\GeV$, as well as for $\ycm>0$ when $60<\mmumu<120\GeV$, and trends in dimuon \pt and \phistar, as mentioned in Section~\ref{sec:samples}.

\subsection{Muon momentum scale and resolution corrections}

The muon momentum scale and resolution are corrected in both data and simulation following the standard CMS procedure described in Ref.~\cite{Bodek:2012id}. These 
corrections have been derived using the \pp data sample at $\sqrts = 13\TeV$ recorded in 2016, with the same detector conditions as the \pPb data set used in the present
analysis. 

In addition, the measurement is unfolded to account for finite momentum resolution. 
No regularisation is found to be needed given the good resolution and modest migrations between the analysis bins, and the maximum likelihood estimate~\cite{Cowan:358560} (obtained from the inversion of the response matrix, derived using simulated NLO \POWHEG samples) is used to obtain the unfolded results.
The effect of the unfolding is less than 1\% in most cases, except for the mass dependence close to the \PZ boson mass peak, where it can amount to up to 15\%.

\subsection{Acceptance and efficiency}

After subtraction of the contributions from different background processes, correction for the muon momentum resolution and scale, 
and unfolding for the detector resolution, the data need to be
corrected for the acceptance and efficiency. The acceptance is defined as the fraction of generated signal events in the full phase space (within the quoted dimuon mass range and $-2.87<\ycm<1.93$) passing the kinematic selection defining the so-called fiducial region: leading muon $\pt>15\GeV$, trailing muon $\pt>10\GeV$, and $\abs{\etalab}<2.4$. Results are presented both with and without this acceptance correction, \ie extrapolated to the full phase space and restricted to the fiducial region, respectively.
The efficiency is the fraction of these events passing all other analysis selection criteria, including trigger selection, muon identification and isolation, and dimuon selection.

The efficiency is also checked in data, using \PZ boson events, with a \textit{tag-and-probe} technique, as described in Ref.~\cite{Khachatryan:2010xn}. The same procedure and corrections are used as in the measurement of $\PW^\pm$ bosons in \pPb collisions~\cite{Sirunyan:2019dox}. The observed differences between the efficiency in data and simulation, estimated separately
for the trigger, identification, and isolation, are accounted for as scale factors on a per-muon basis that are applied to the simulated events. These corrections are applied both in the efficiency estimation and in the construction of the background templates described in Section~\ref{sec:bkg}. 
When both muons in the event have $\pt>15\GeV$, they can both pass the single-muon trigger used in this data analysis, and the scale factor is computed from the product of inefficiencies.
For the muon and central track reconstruction, the data and MC simulation are found to give 
comparable efficiencies ($> 99.9\%$) and therefore no scale factor is applied for these two components of the efficiency.

\subsection{Final-state radiation effects}
\label{sec:FSR}

Muons may undergo final-state radiation before being measured in the CMS detector, biasing their momentum and shifting the dimuon mass to lower values. We unfold the measured
distributions, after efficiency correction (as well as acceptance, if applicable), to the ``pre-FSR'' quantities, used for the presentation of our results and defined from a ``dressed lepton'' definition~\cite{CMS:2014jea}. Generator-level muon four-momenta are recalculated by adding the four-momenta of all generated photons found inside a cone of radius $\Delta R = 0.1$ around the muon.
Again the response matrices for this unfolding procedure, derived using simulated NLO \POWHEG samples, are found to be close to diagonal, thus no regularisation is needed in the unfolding.

\subsection{Systematic uncertainties}

Several sources of systematic uncertainties are evaluated. They are estimated in each bin of the measured distributions and added in quadrature. The list of systematic uncertainties is summarised in \tab{tab:syst} and details of the estimation of each source are given below.

Theoretical uncertainties have an impact on the acceptance and efficiency. The renormalisation and factorisation scales have been varied from half to twice their nominal value (set to the 
dimuon mass), and the envelope of the variations, excluding combinations where both scales are varied in opposite directions, is taken as an uncertainty. In addition, the strong coupling constant value is varied by 0.0015 from its default value,
$\alpS (m_{\PZ}) = 0.118$, as recommended by PDF4LHC~\cite{Butterworth:2015oua}. The CT14 and EPPS16 uncertainties are also included, estimated with \textsc{LHAPDF6}~\cite{lhapdf6} using the PDF4LHC recommendations for Hessian (n)PDF sets~\cite{Butterworth:2015oua}. 
Finally, the full difference between the acceptance and efficiency obtained with and without the \PZ boson \pt reweighting is
considered as a systematic uncertainty. The impact of these uncertainties is less than 1\% on the efficiency, but up to 10\% on the acceptance for low dimuon masses.

We also include uncertainties stemming from the estimation of the efficiencies from data. The statistical component coming from the limited \PZ boson sample available is treated as 
a systematic uncertainty in this analysis. We also consider systematic effects associated with the choice of function used to model the \pt behaviour of the efficiencies, the dimuon mass fitting
procedure to the \PZ boson peak in the extraction of the efficiencies, a possible data-to-simulation difference in the muon reconstruction efficiency, and the effect of the mismodelling
in simulation of the event activity and for additional interactions per bunch crossing. The magnitude of these uncertainties ranges from 1 to 5\% at low dimuon mass.

Regarding the estimation of EW backgrounds with the \emu method, the statistical uncertainty in the correction factors is included as a systematic uncertainty, 
as well as the effect of varying the \ttbar cross section by its uncertainty, 18\%~\cite{Sirunyan:2017xku}, the uncertainty in the transfer factor for the heavy-flavour contribution, and the difference between the data and simulation in the \emu distributions. 
The systematic uncertainty in the multijet and {\PW}+jets backgrounds, related to the misidentification rate method, receives several contributions. The statistical uncertainty in the templates derived from data is accounted for, and combined with the full difference between the nominal estimation and an alternative method (based on a different sideband in data, using same-sign dimuon events). The residual nonclosure in the same-sign data sample, as well as its statistical uncertainty, are also both added in quadrature to the other uncertainties related to the misidentification rate method. The total systematic uncertainty in the background estimate, dominated by the residual nonclosure in most bins, ranges from less than 0.5 to 15\% (for large dimuon \pt).

A different reweighting of the event activity in simulated samples is derived, as a function of the number of offline tracks reconstructed with $\abs{\etalab}<2.5$ instead of the nominal correction using the
total energy deposited in the HF calorimeters, which modifies the efficiency and the background estimation. The observed difference in the measurements, which is less than 1\% in most bins, is taken as a systematic uncertainty.

Uncertainties in the muon momentum scale and resolution corrections have been evaluated, based on the 2016 \pp data sample at $\sqrts = 13\TeV$, from which they are derived. These uncertainties, about 1\% or less,
arise from the limited data sample size available and variations in the method and its assumptions. 

Response matrices used in the muon momentum scale and FSR unfoldings have been re-calculated using the first and second parts of the run alone (accounting for statistical uncertainties in simulation), and using the \PYQUEN generator v1.5.1~\cite{Lokhtin:2005px} instead of \POWHEG (for a conservative estimation of the model dependence). Differences in the unfolded results, which are up to 2\%, are taken into account as a systematic uncertainty.

Finally, the uncertainty in the integrated luminosity measurement is 3.5\%~\cite{LUM17002}.

\begin{table}[htb]
  \topcaption{
 \label{tab:syst}
 Range of systematic uncertainties in percentage of the cross section, given separately for $15<\mmumu<60$ and $60<\mmumu<120\GeV$. Systematic uncertainties for the three mass bins above 120\GeV fall in the range given for $15<\mmumu<60\GeV$. For the theoretical component of acceptance and efficiency, the systematic uncertainty related to efficiency alone (for fiducial cross sections) is given between parentheses.
 }
 \centering
\begin{tabular}{lcc}
 Source of uncertainty & $15<\mmumu<60\GeV$ & $60<\mmumu<120\GeV$ \\
 \hline
 Event activity reweighting & $<$3\% & $<$1\% \\
 Muon momentum & $<$1\% & $<$3\% \\
 Data-driven efficiencies & 1--5\% & 1--4\% \\
 Acceptance and efficiency (MC stat.) & $<$4\% & $<$4\% \\
 Background estimation & 2--15\% & 0.1--3\% \\
 Acceptance and efficiency (theory) & 1--10\% ($<$1\%) & $<$1\% ($<$1\%) \\
 Unfolding: detector resolution & $<$2\% & $<$2\% \\
 Unfolding: FSR & $<$1\% & $<$1\% \\
 \\
 Total & 6--15\% & 1--12\%
\end{tabular}
\end{table}

\begin{figure}[htbp] 
\centering
\includegraphics[width=0.7\textwidth]{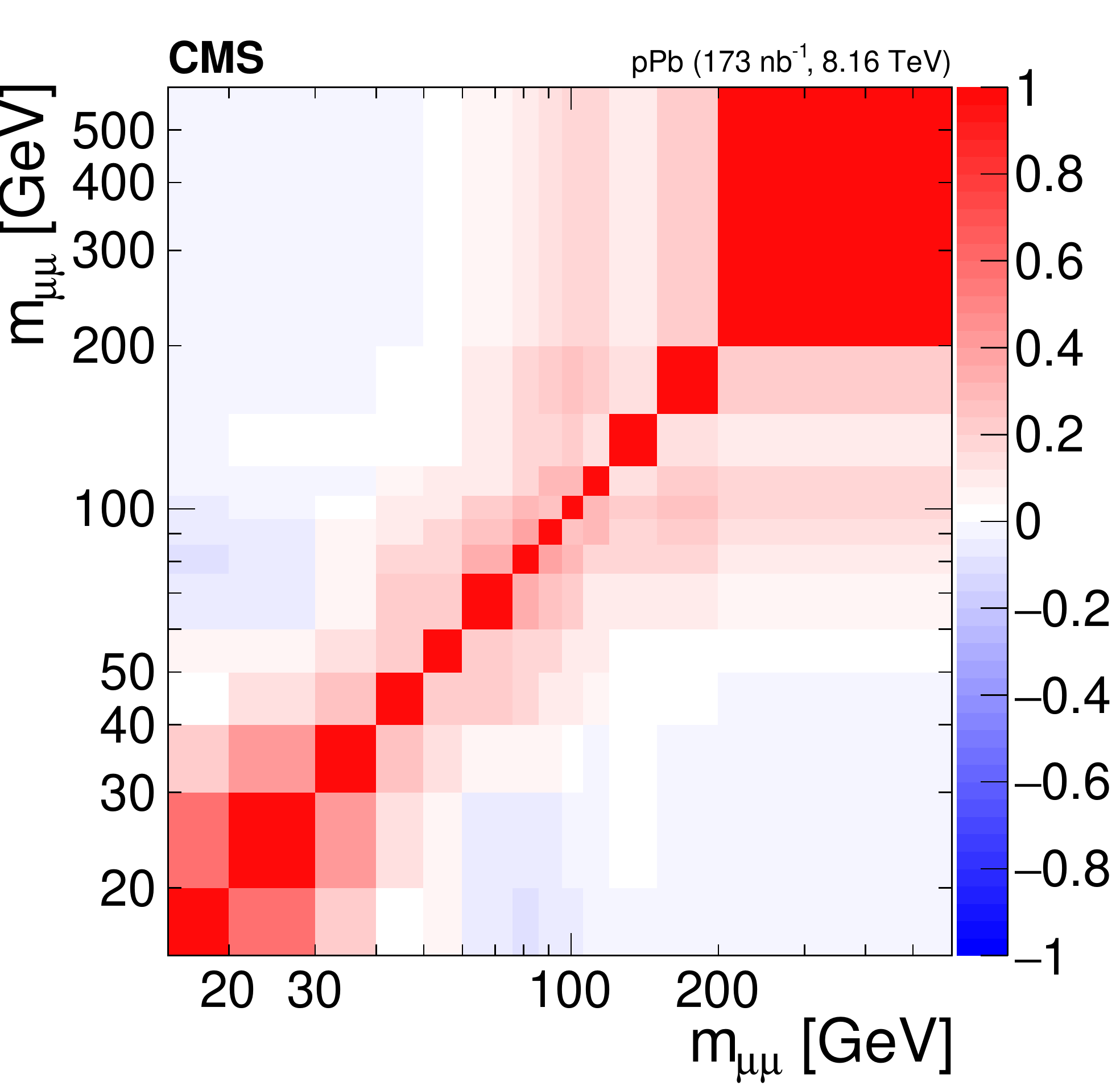}

\includegraphics[width=0.49\textwidth]{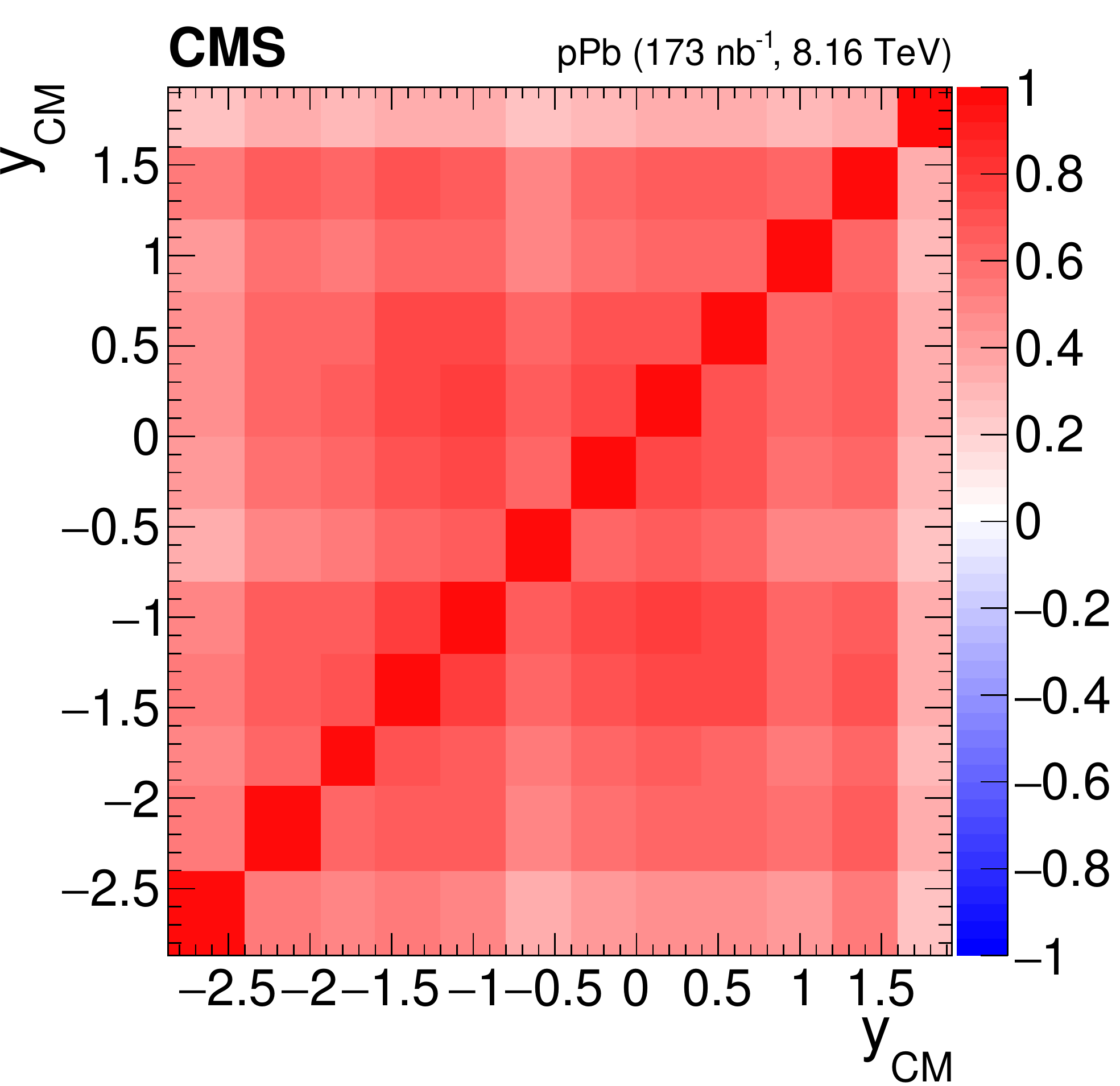} 
\includegraphics[width=0.49\textwidth]{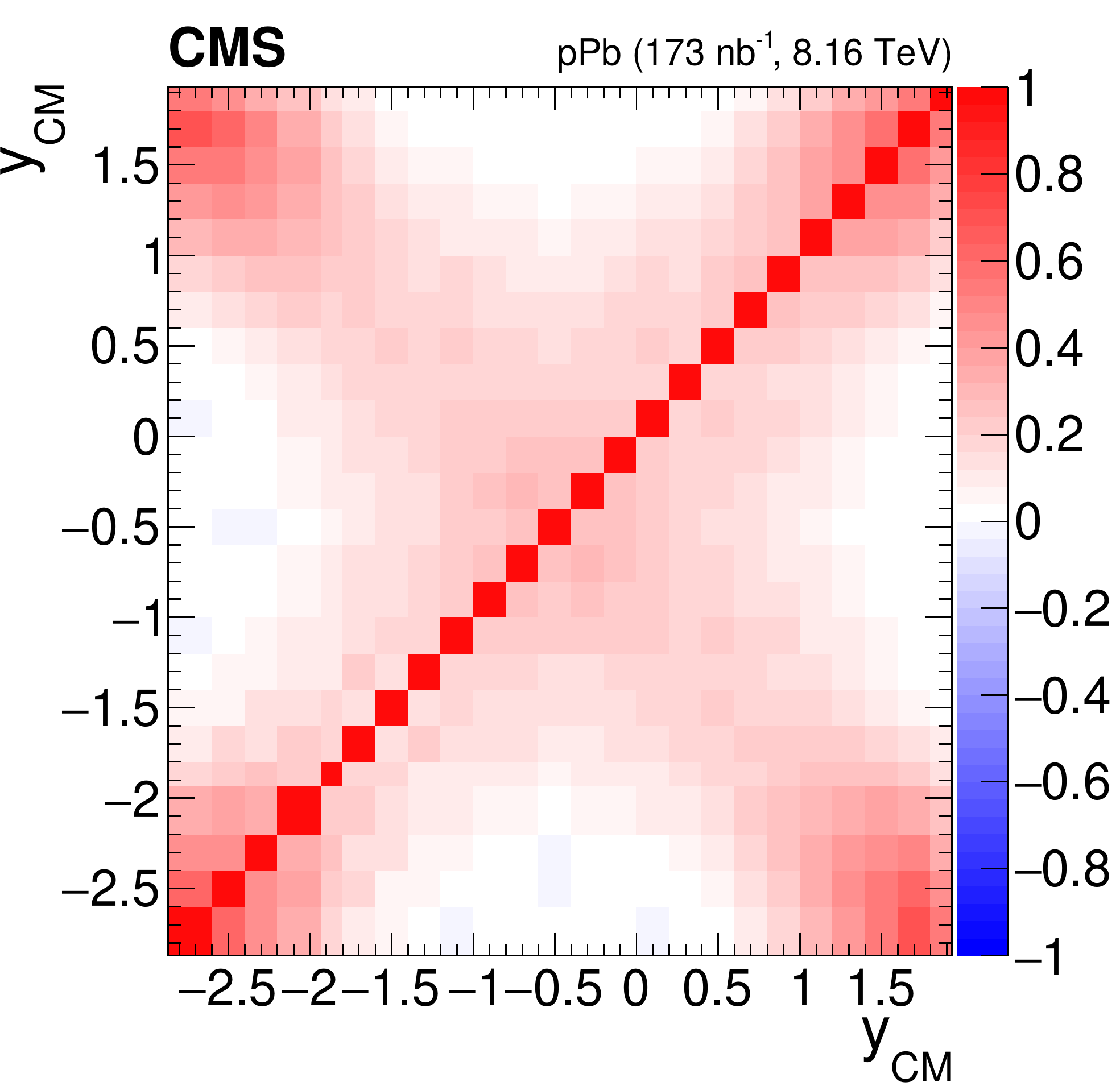} 
\caption{
\label{fig:cormat_mass}
Correlation matrix for the systematic uncertainties, excluding integrated luminosity,
   as a function of the dimuon invariant mass (upper) and rapidity in the centre-of-mass frame for $15<\mmumu<60\GeV$ (lower left) and $60<\mmumu<120\GeV$ (lower right).
}
\end{figure}

\begin{figure}[htbp] 
 \centering

\includegraphics[width=0.49\textwidth]{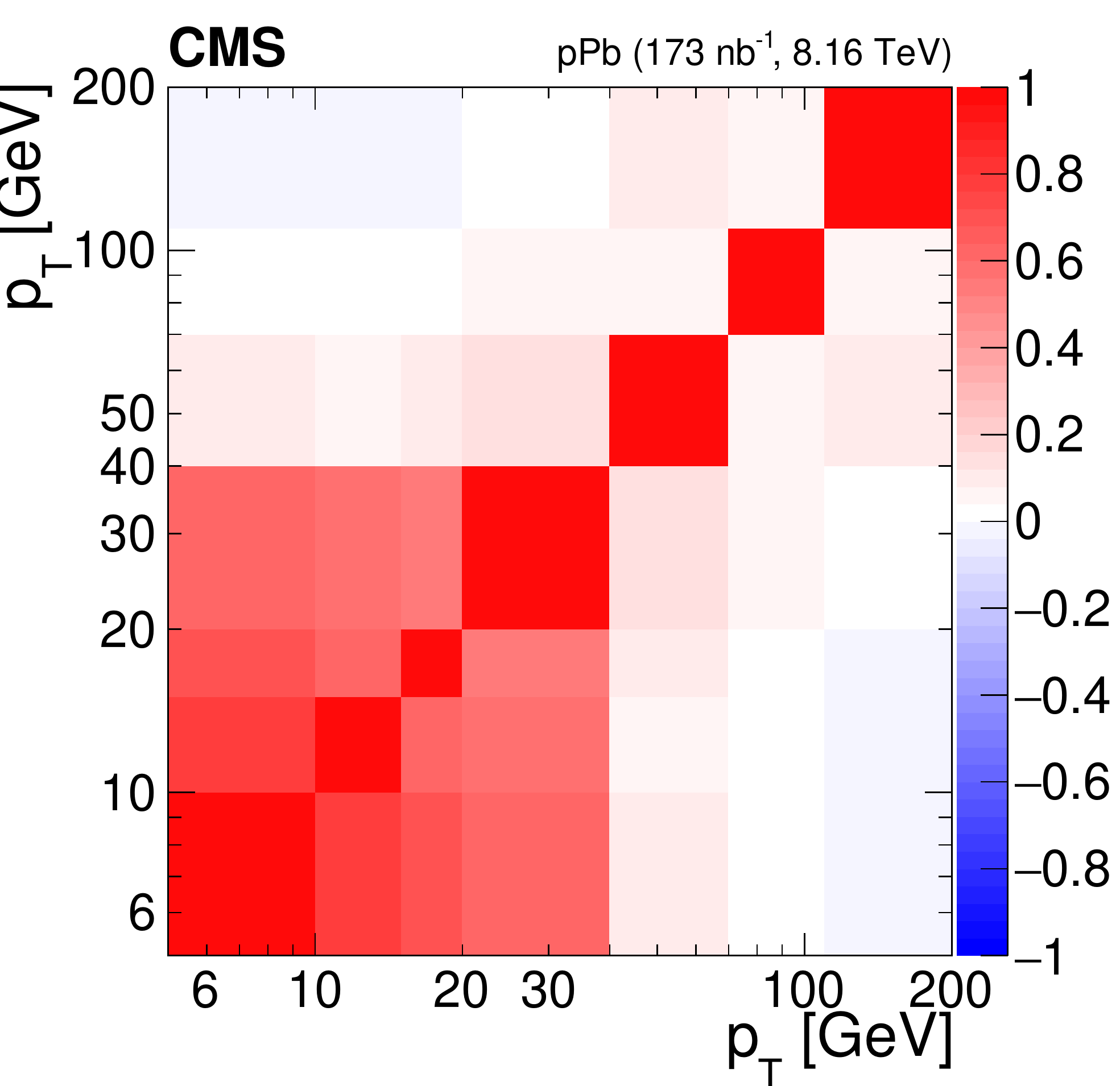} 
\includegraphics[width=0.49\textwidth]{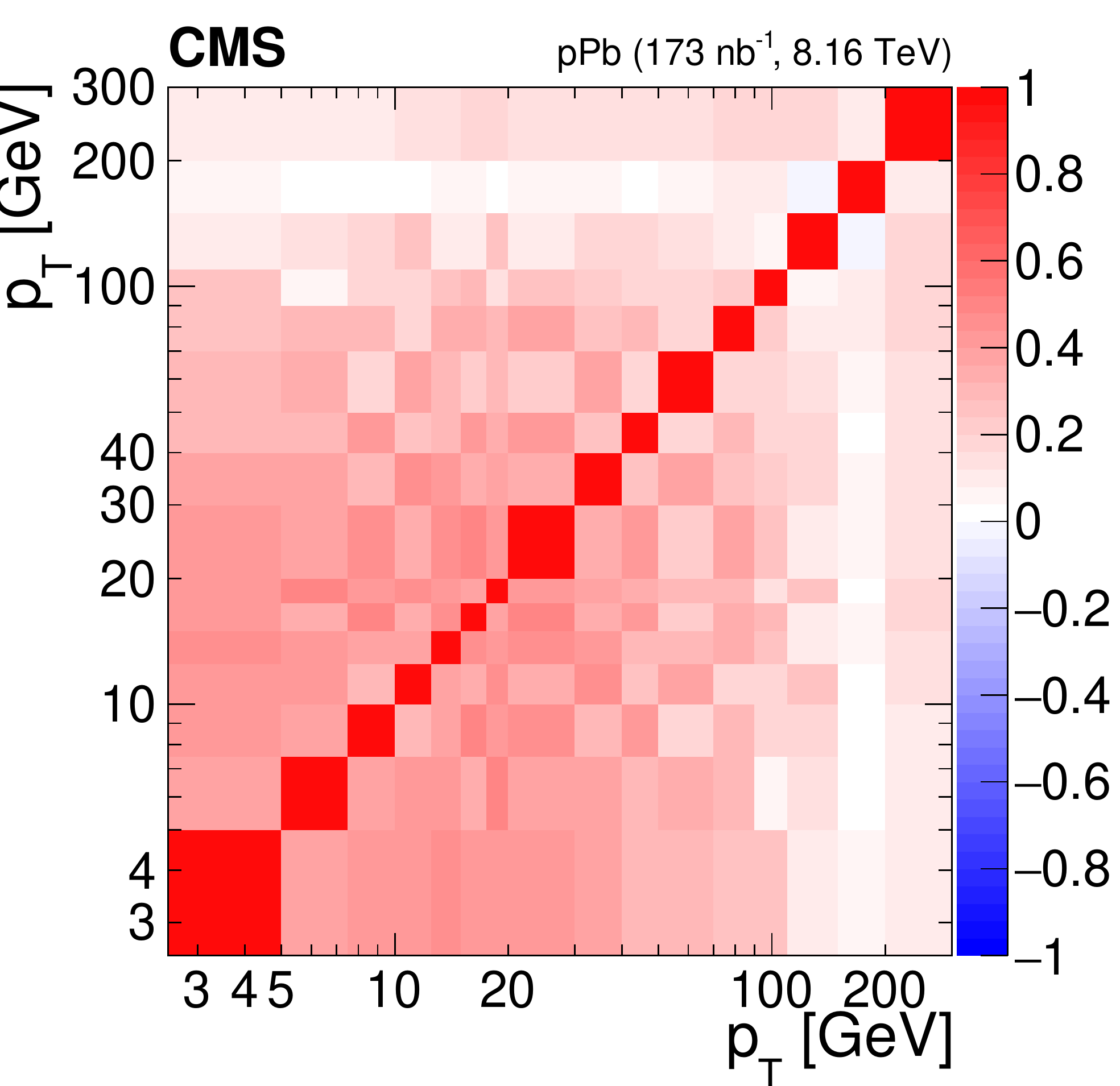}

\includegraphics[width=0.49\textwidth]{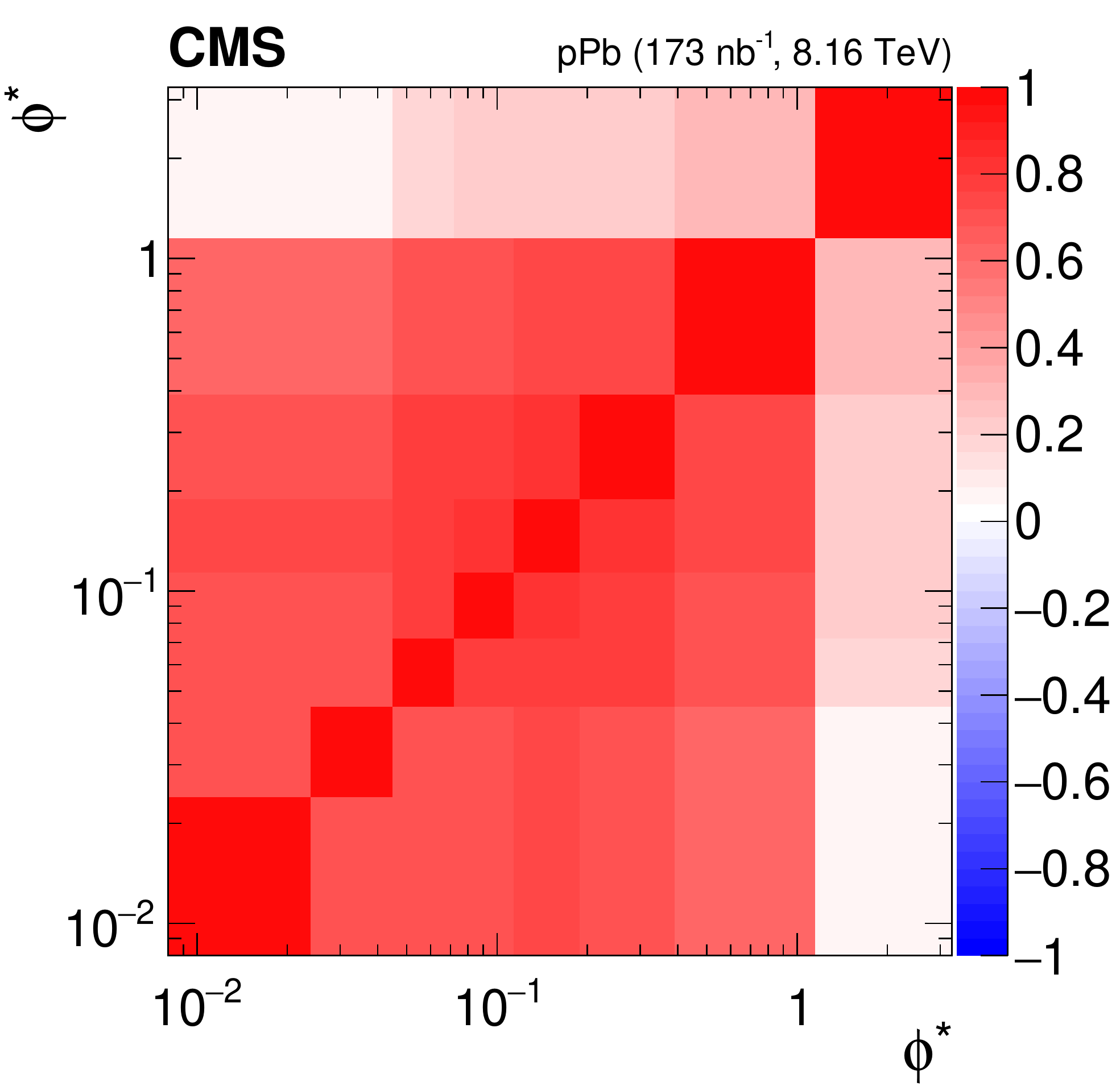} 
\includegraphics[width=0.49\textwidth]{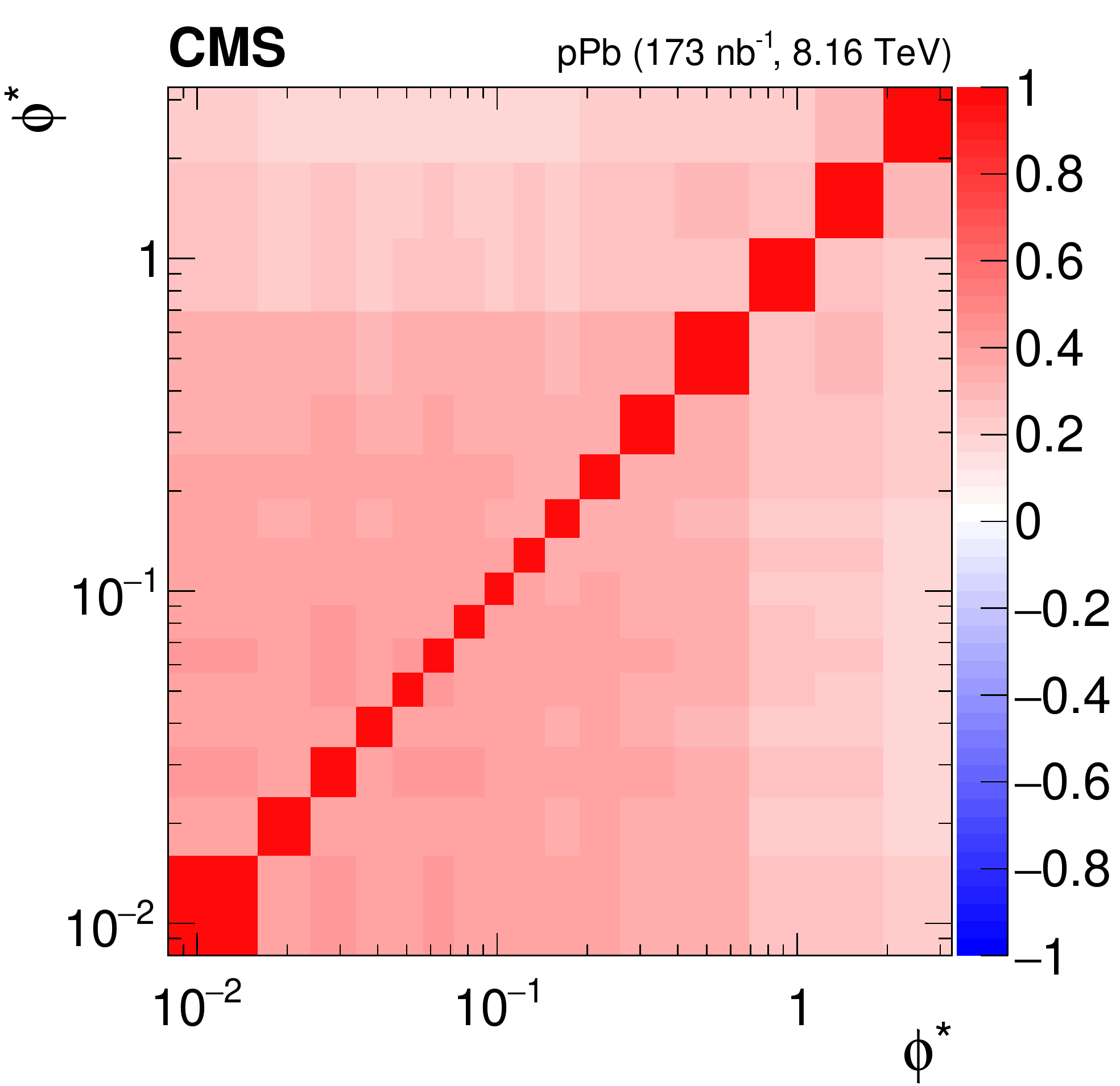} 
\caption{
\label{fig:cormat_other}
Correlation matrices for the systematic uncertainties, excluding integrated luminosity,
as functions of \pt (upper row) and \phistar (lower row), for $15<\mmumu<60\GeV$ (left) and $60<\mmumu<120\GeV$ (right).
}
\end{figure}

Correlations across bins of these uncertainties have also been evaluated. Theoretical uncertainties are assumed to be fully correlated, with the exception of the nPDF uncertainty, whose correlation is calculated using 
the CTEQ prescription for Hessian sets~\cite{Lai:2010vv}. Systematic uncertainties in the efficiency scale factors obtained from control samples in data are assumed to be uncorrelated, since they could
have different effects in different kinematic regions, while statistical correlations between the scale factors derived in the same region of the detector are accounted for. No correlation 
is assumed for the uncertainties related to the background estimation. Uncertainties related to the HF energy reweighting, unfolding, and integrated luminosity are treated as fully correlated between the bins and measurements, as well as each of
the sources of uncertainty in the muon momentum scale and resolution corrections. The correlation matrices for systematic uncertainties are shown in Figs.~\ref{fig:cormat_mass} and ~\ref{fig:cormat_other}, excluding the fully correlated integrated luminosity uncertainty for clarity. They are derived from the total covariance matrix, obtained from the sum of the covariance matrices for the individual sources, assuming the correlations above. For a given variable, the difference between the matrices in the two mass selections can be explained by the background uncertainty, which is one of the dominant systematic uncertainties for $15<\mmumu<60\GeV$ but negligible most of the time for $60<\mmumu<120\GeV$, except at large \pt or \phistar.
Muon efficiency uncertainties, treated as a function of $\abs{\etalab}$, induce a weak anticorrelation visible in systematic uncertainties as a function of rapidity, especially visible in the $60<\mmumu<120\GeV$ region where they are the dominant systematic uncertainty.

\section{Results and discussion}
\label{sec:results}

\begin{figure}[htbp]
\centering
\includegraphics[width=0.7\textwidth]{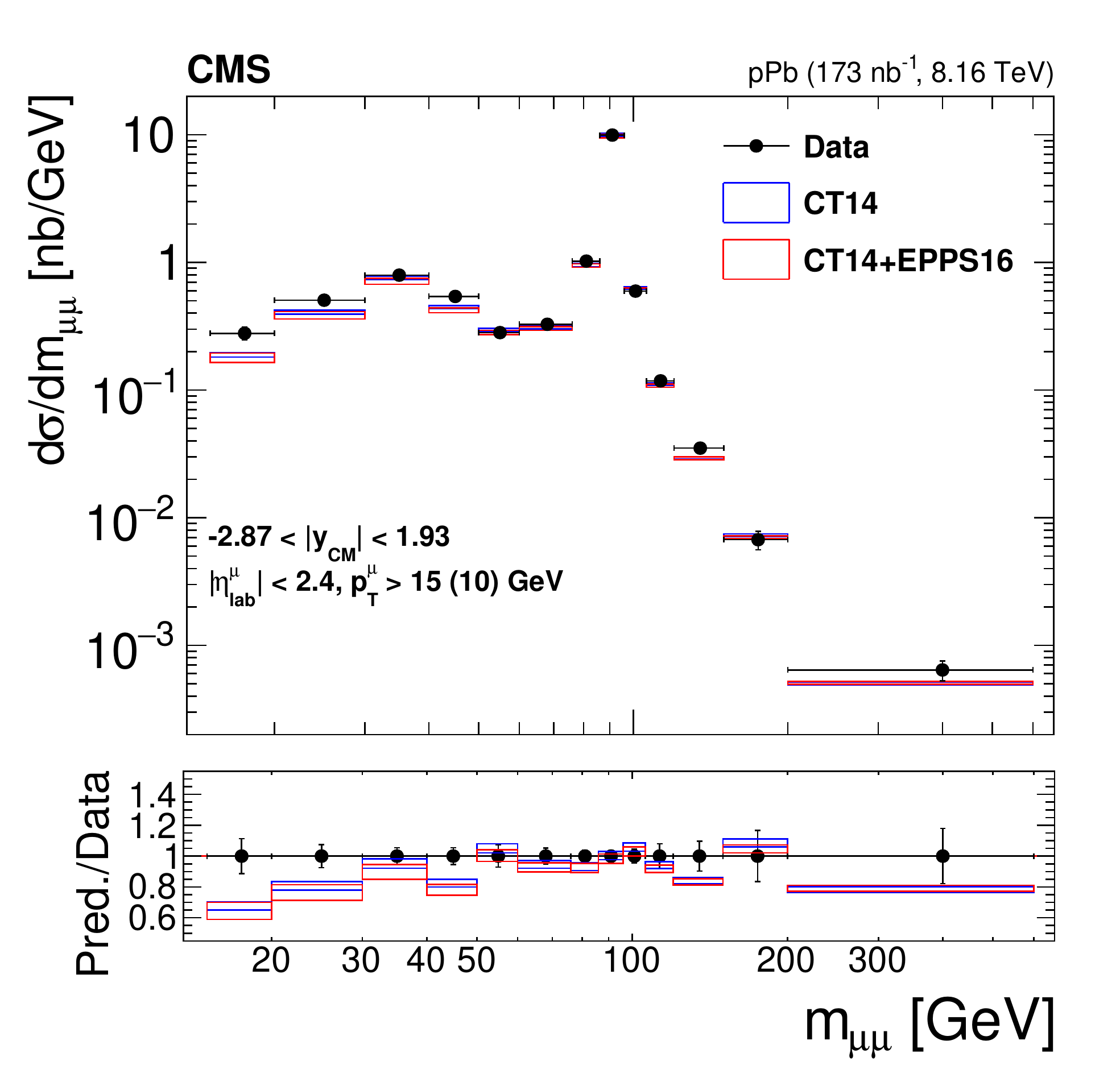}

\includegraphics[width=0.49\textwidth]{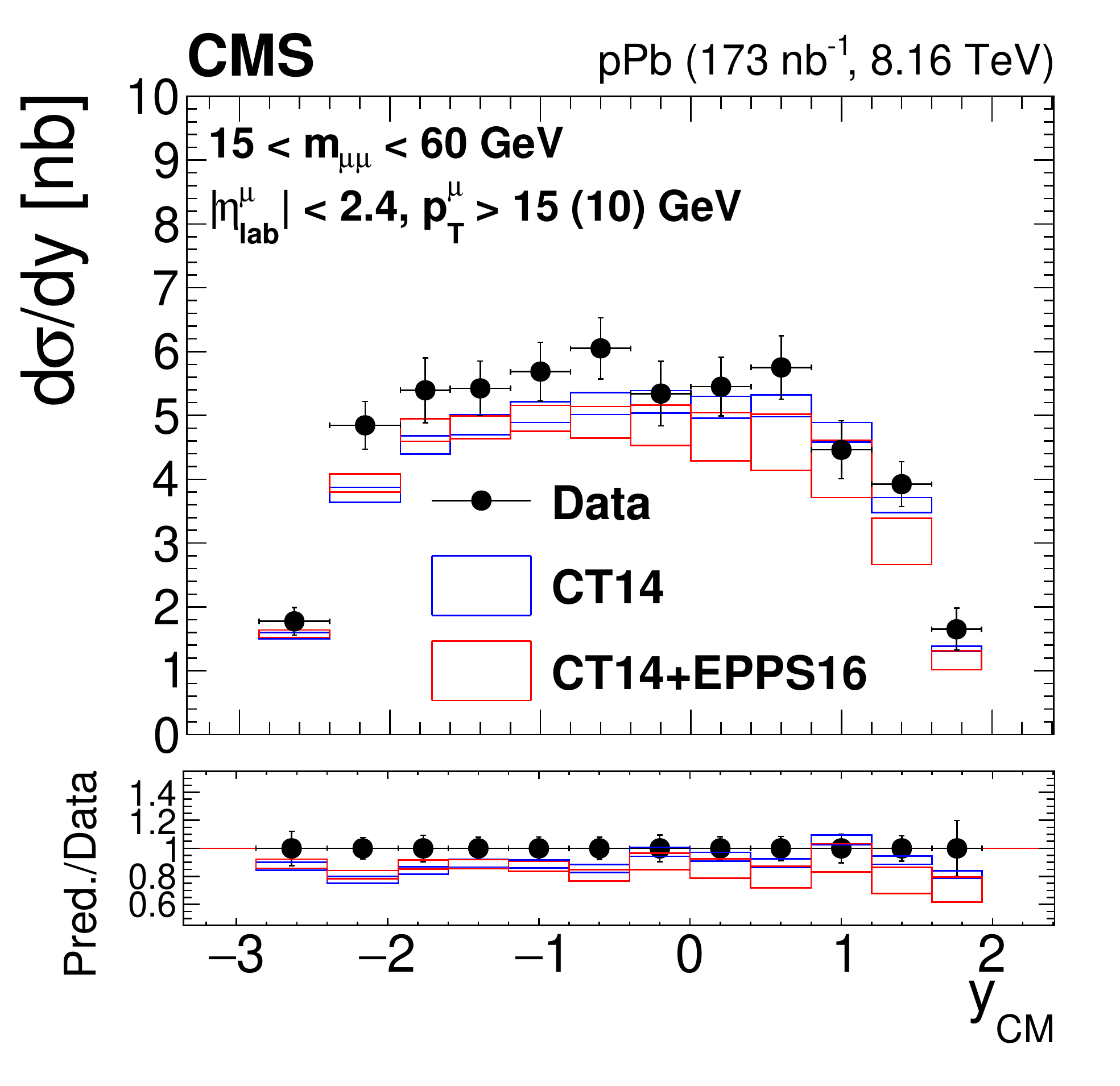} 
\includegraphics[width=0.49\textwidth]{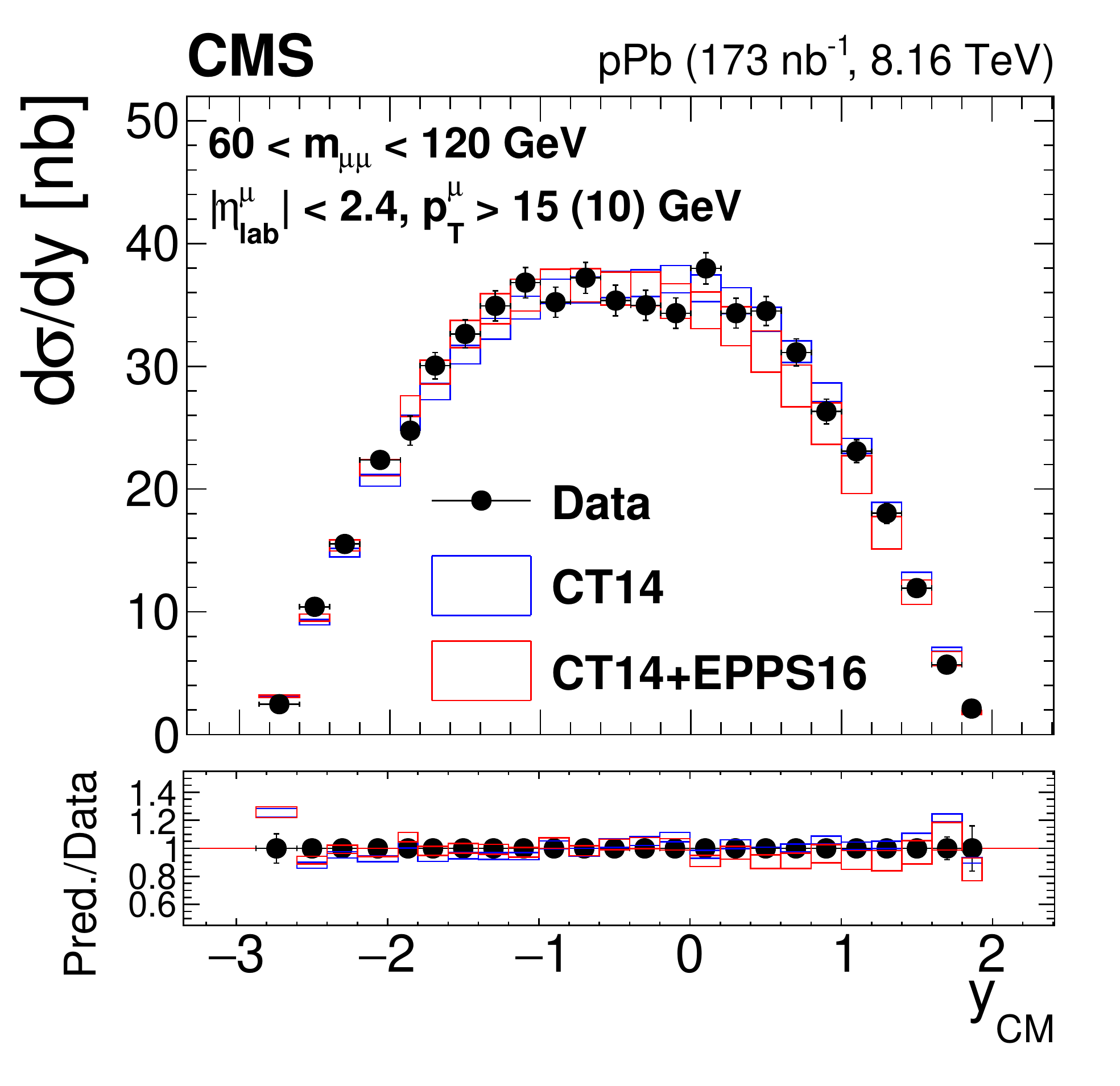} 
\caption{
\label{fig:xsec_noacc_mass}
Differential fiducial cross section (without the acceptance correction) for the DY process measured in the muon channel,
   as a function of the dimuon invariant mass (upper) and rapidity in the centre-of-mass frame for $15<\mmumu<60\GeV$ (lower left) and $60<\mmumu<120\GeV$ (lower right).
The error bars on the data represent the quadratic sum of the statistical and systematic uncertainties. Theory predictions from the \POWHEG NLO generator
are also shown, using CT14 (blue) or CT14+EPPS16 (red). The boxes show the 68\% confidence level (n)PDF uncertainty on these predictions. The ratios of predictions over data are shown in the lower panels, where the data and (n)PDF uncertainties are shown separately, as error bars around one and as coloured boxes, respectively.
}
\end{figure}

\begin{figure}[htbp]
\centering

\includegraphics[width=0.49\textwidth]{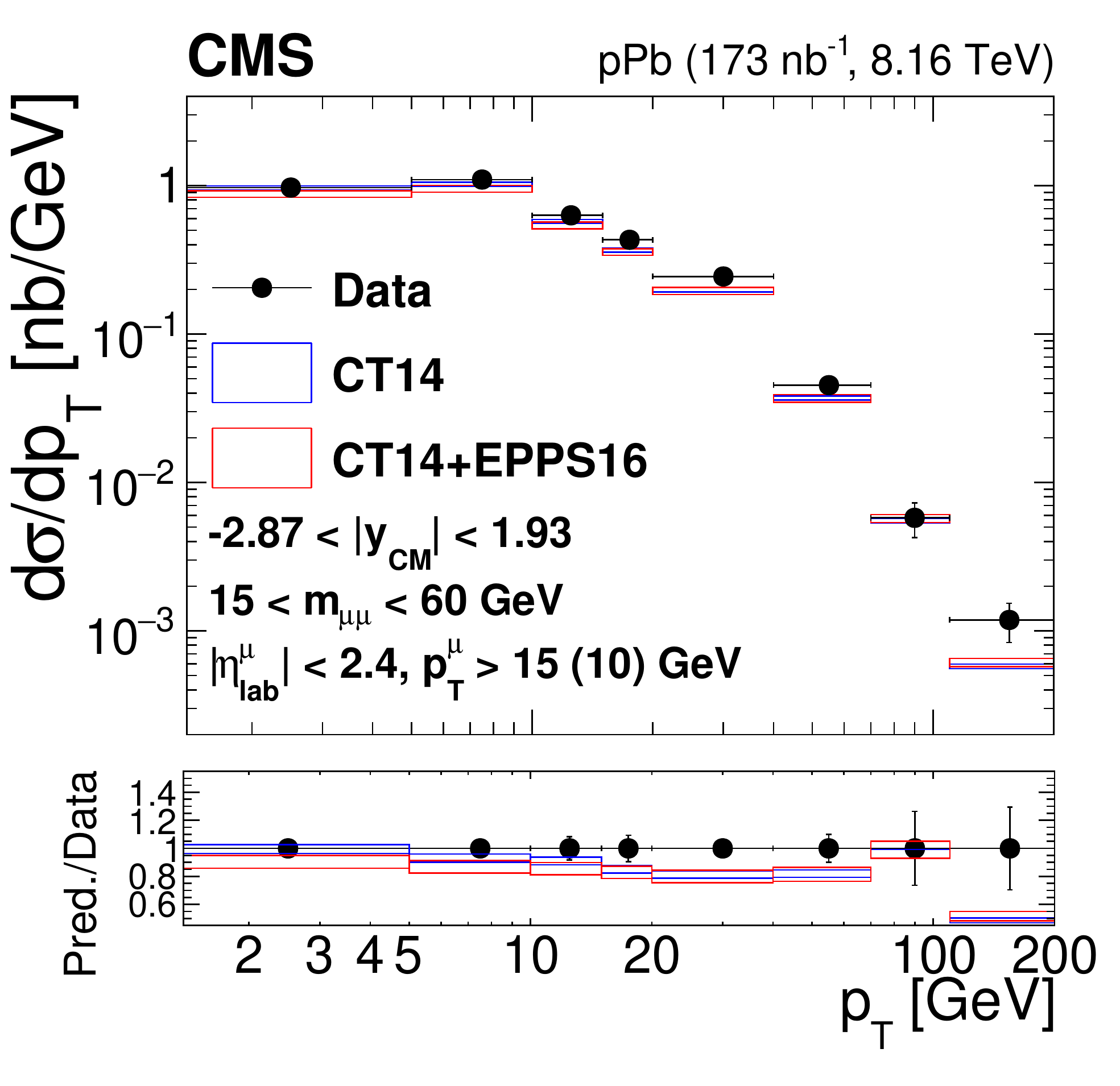} 
\includegraphics[width=0.49\textwidth]{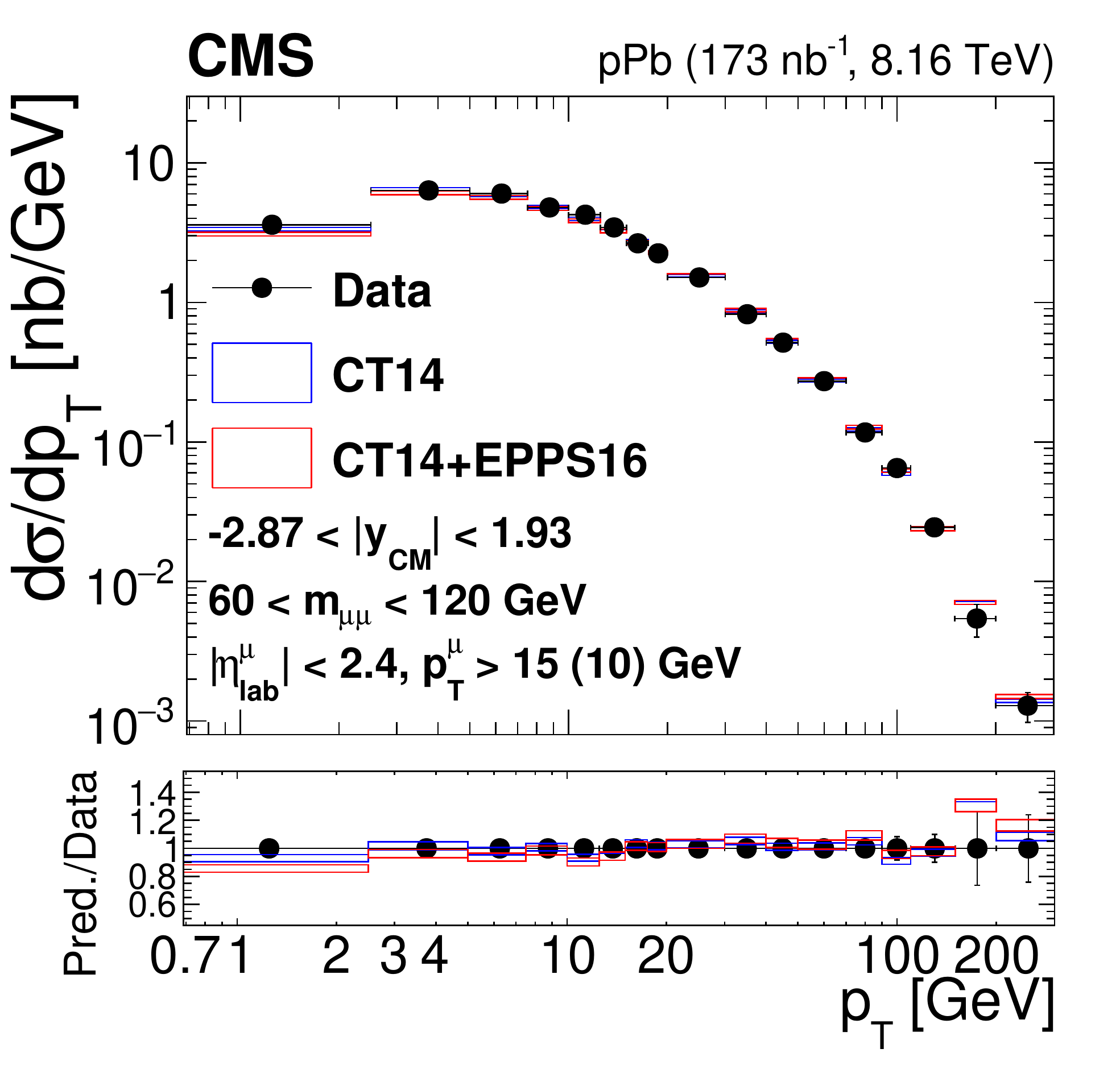}

\includegraphics[width=0.49\textwidth]{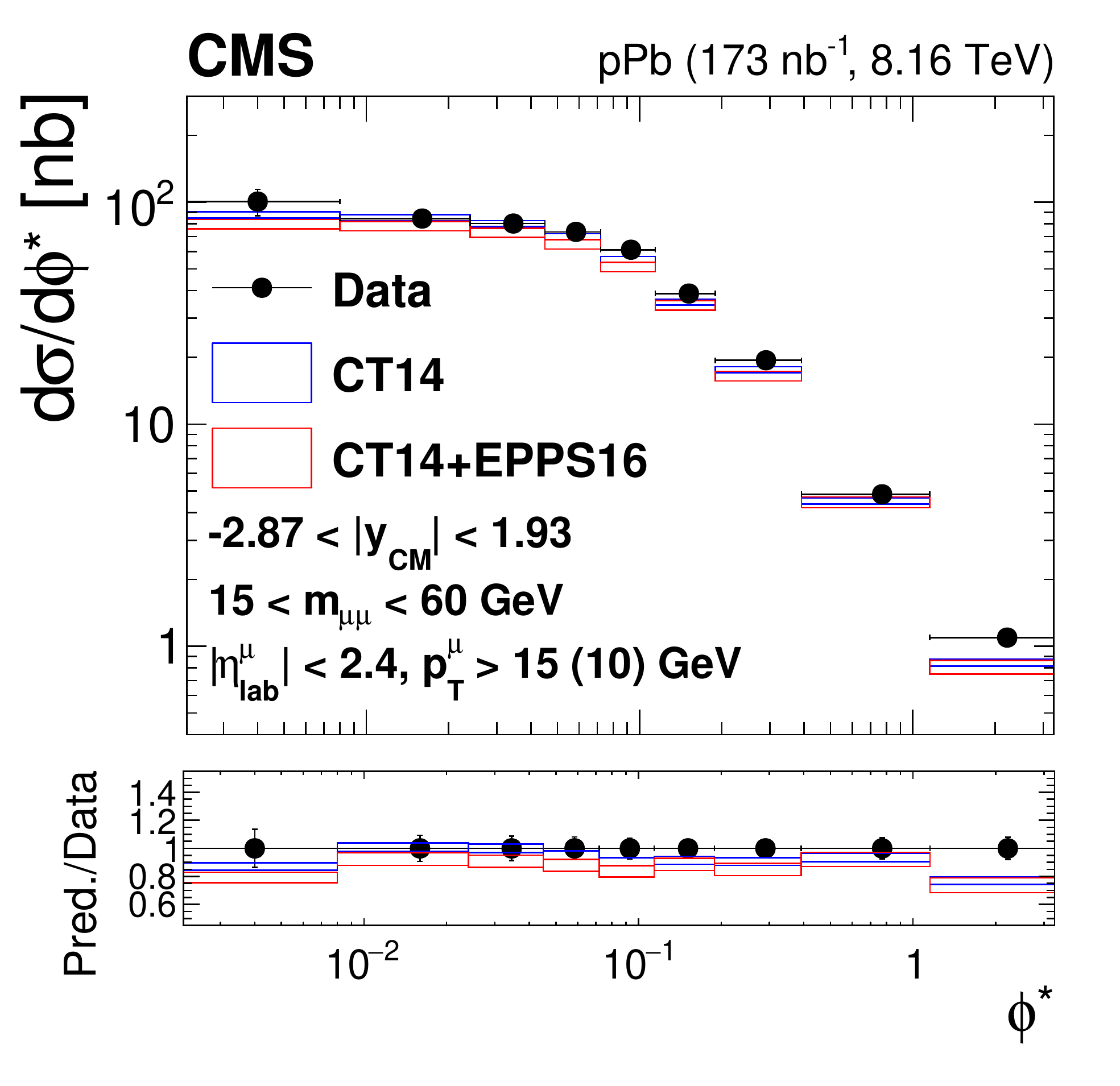} 
\includegraphics[width=0.49\textwidth]{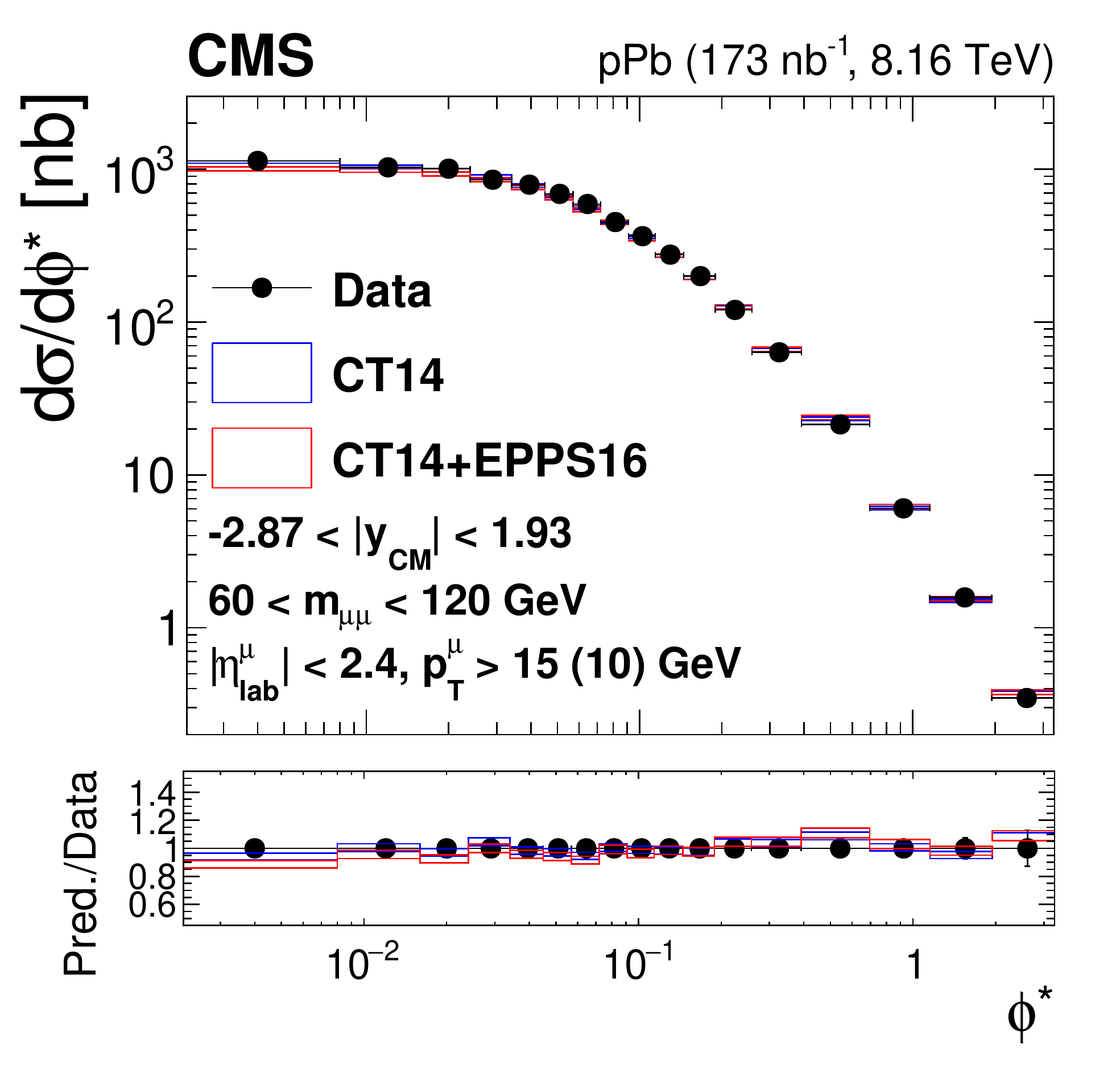} 
\caption{
\label{fig:xsec_noacc_other}
Differential fiducial cross sections (without the acceptance correction) for the DY process measured in the muon channel,
as functions of \pt (upper row) and \phistar (lower row), for $15<\mmumu<60\GeV$ (left) and $60<\mmumu<120\GeV$ (right).
The first bin of the \pt and \phistar measurements starts at 0.
The error bars on the data represent the quadratic sum of the statistical and systematic uncertainties. Theory predictions from the \POWHEG NLO generator
are also shown, using CT14 (blue) or CT14+EPPS16 (red). The boxes show the 68\% confidence level (n)PDF uncertainty on these predictions. The ratios of predictions over data are shown in the lower panels, where the data and (n)PDF uncertainties are shown separately, as error bars around one and as coloured boxes, respectively.
}
\end{figure}

\begin{figure}[htbp]
\centering
\includegraphics[width=0.7\textwidth]{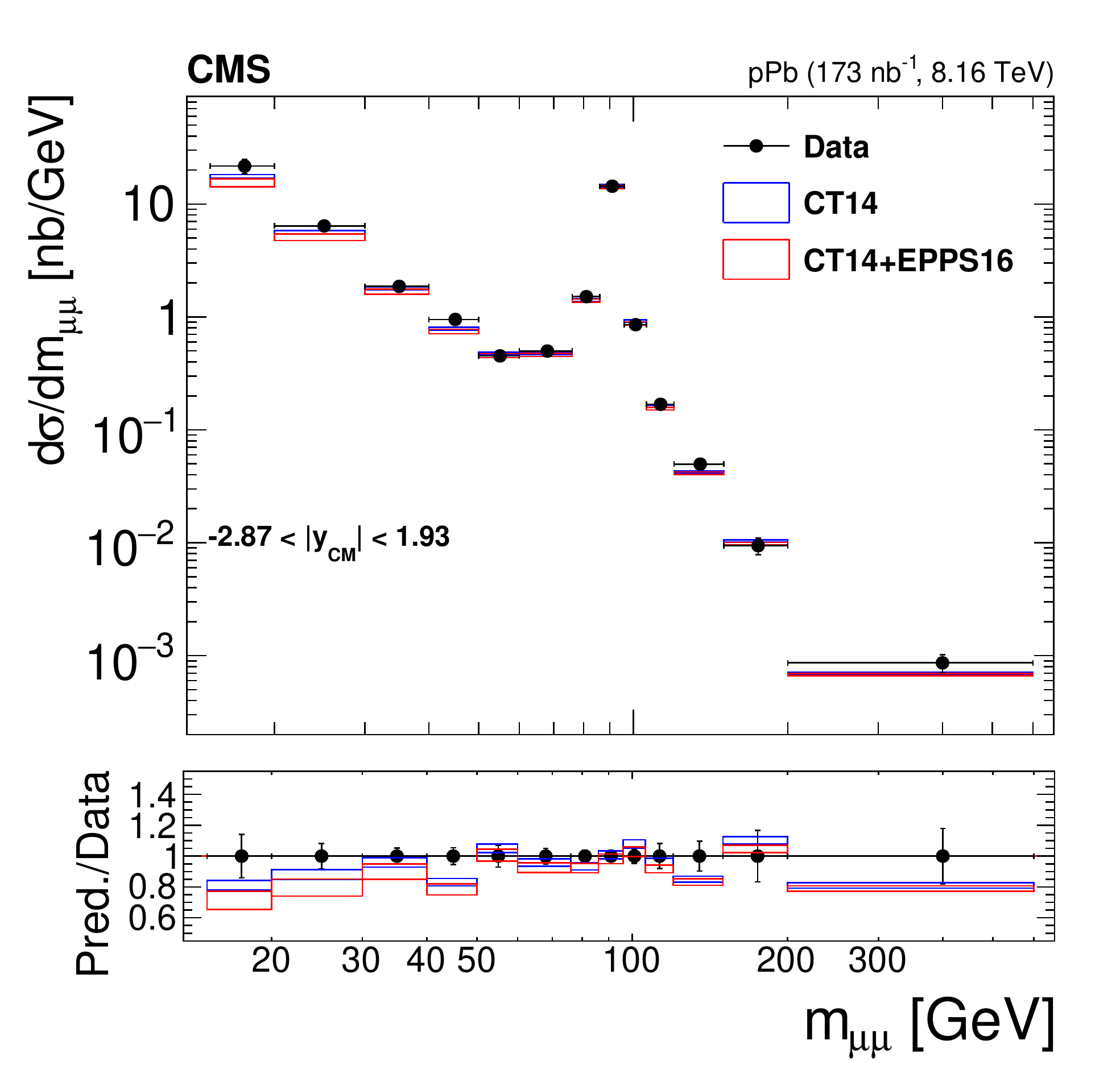}

\includegraphics[width=0.49\textwidth]{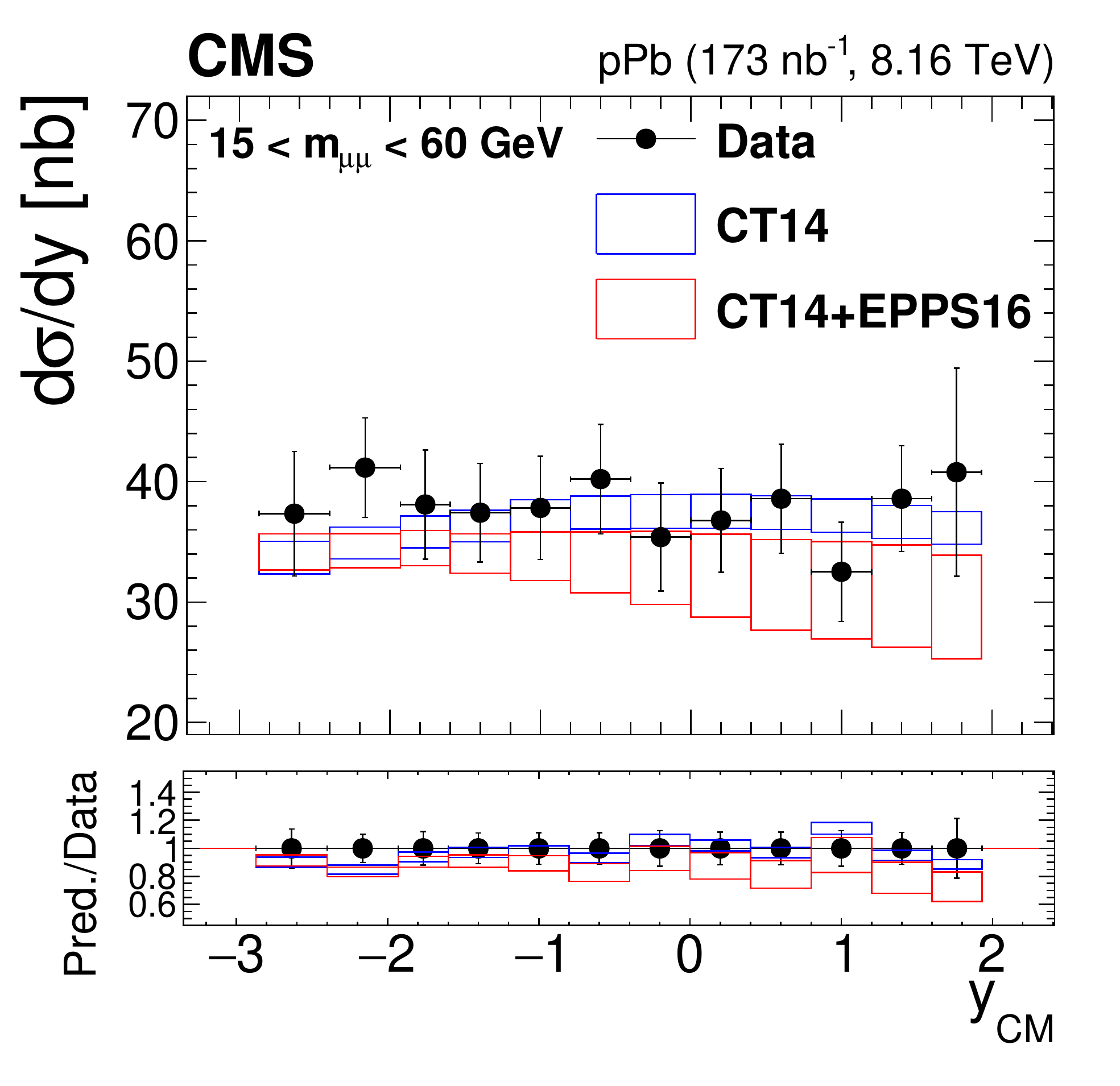} 
\includegraphics[width=0.49\textwidth]{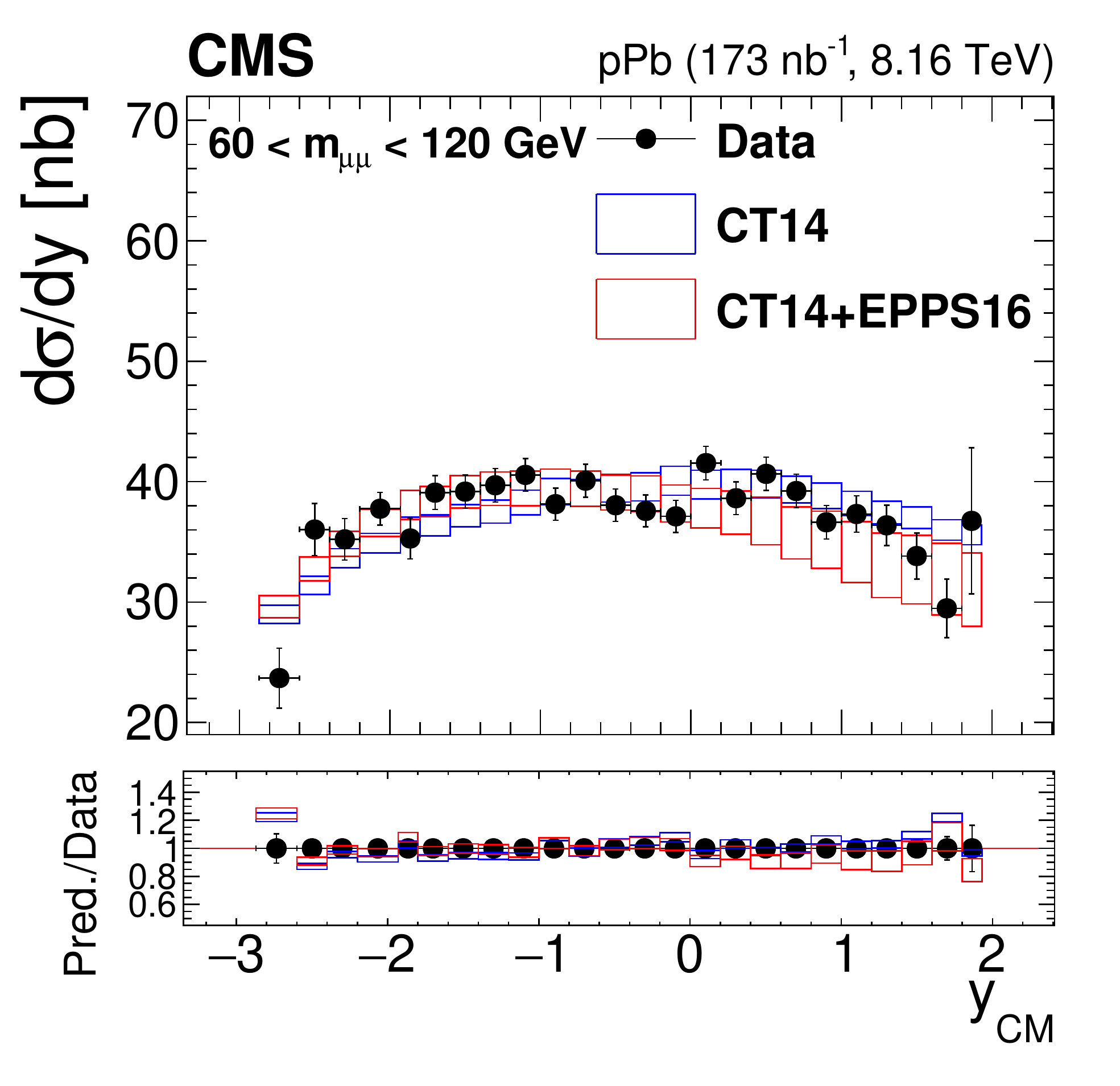} 
\caption{
\label{fig:xsec_mass}
Differential cross section for the DY process measured in the muon channel,
   as a function of the dimuon invariant mass (upper) and rapidity in the centre-of-mass frame for $15<\mmumu<60\GeV$ (lower left) and $60<\mmumu<120\GeV$ (lower right).
The error bars on the data represent the quadratic sum of the statistical and systematic uncertainties. Theory predictions from the \POWHEG NLO generator
are also shown, using CT14 (blue) or CT14+EPPS16 (red). The boxes show the 68\% confidence level (n)PDF uncertainty on these predictions. The ratios of predictions over data are shown in the lower panels, where the data and (n)PDF uncertainties are shown separately, as error bars around one and as coloured boxes, respectively.
}
\end{figure}

\begin{figure}[htbp]
\centering

\includegraphics[width=0.49\textwidth]{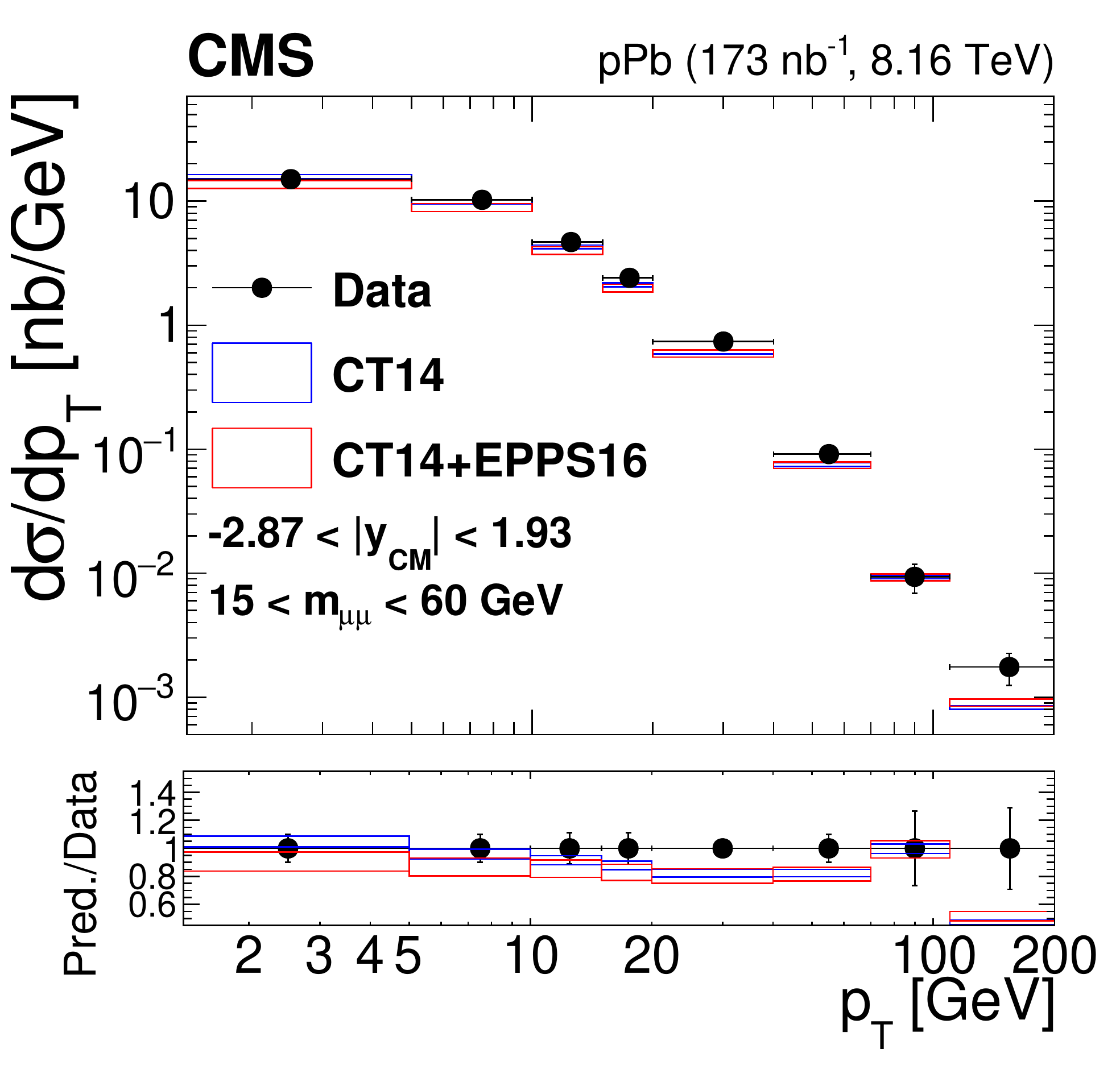} 
\includegraphics[width=0.49\textwidth]{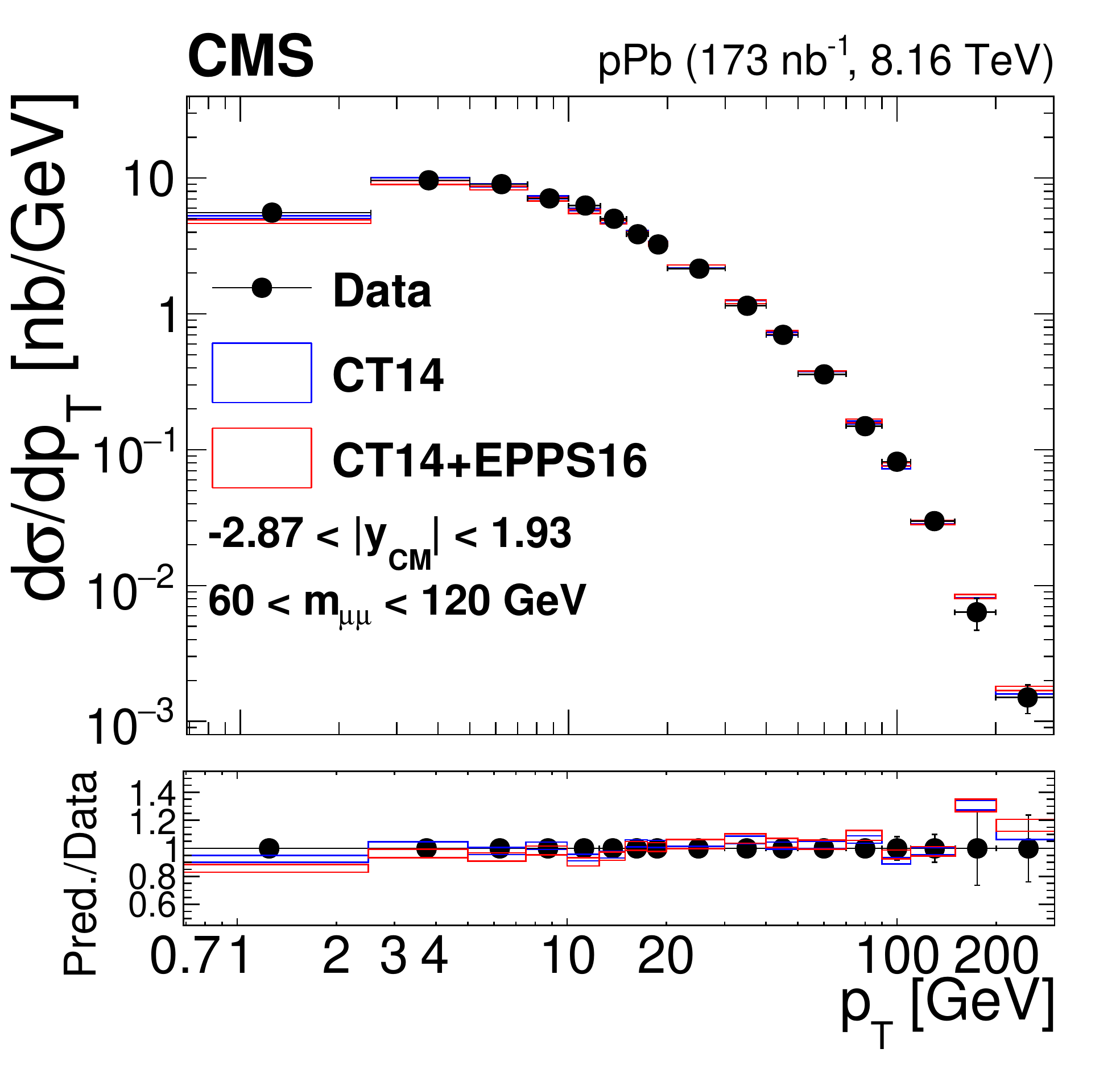}

\includegraphics[width=0.49\textwidth]{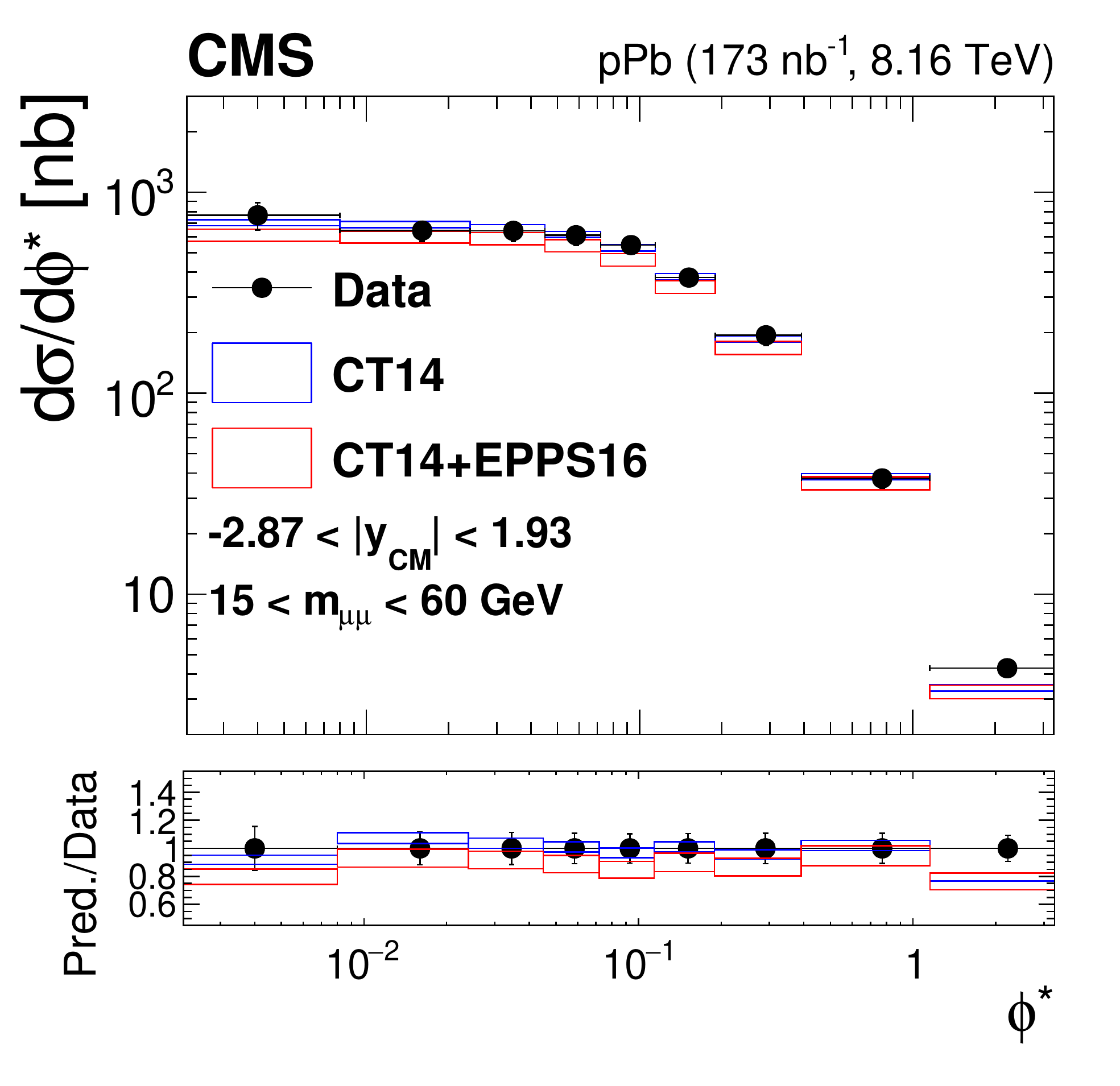} 
\includegraphics[width=0.49\textwidth]{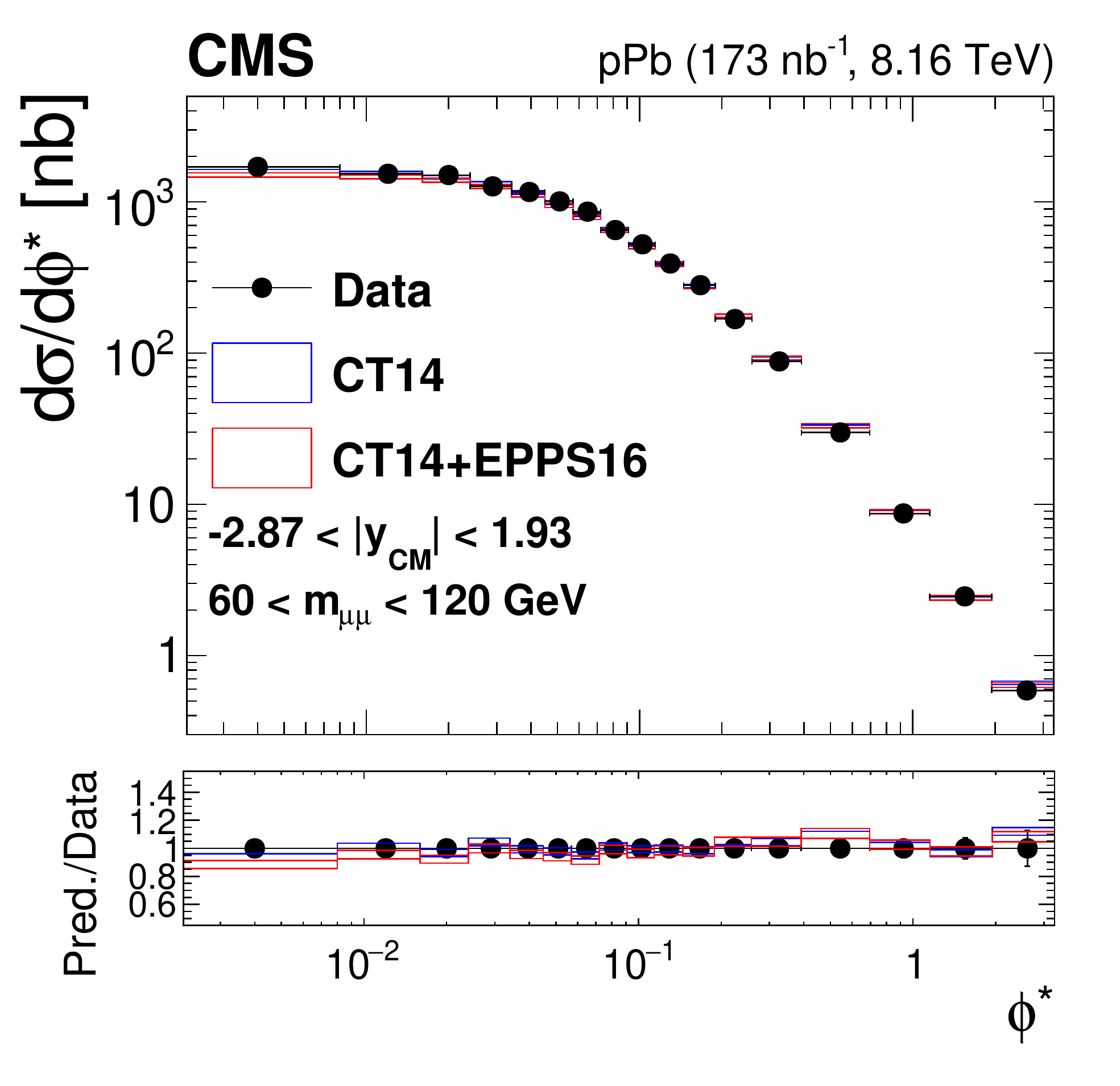} 
\caption{
\label{fig:xsec_other}
Differential cross sections for the DY process measured in the muon channel,
as functions of \pt (upper row) and \phistar (lower row), for $15<\mmumu<60\GeV$ (left) and $60<\mmumu<120\GeV$ (right).
The first bin of the \pt and \phistar measurements starts at 0.
The error bars on the data represent the quadratic sum of the statistical and systematic uncertainties. Theory predictions from the \POWHEG NLO generator
are also shown, using CT14 (blue) or CT14+EPPS16 (red). The boxes show the 68\% confidence level (n)PDF uncertainty on these predictions. The ratios of predictions over data are shown in the lower panels, where the data and (n)PDF uncertainties are shown separately, as error bars around one and as coloured boxes, respectively.
}
\end{figure}

Fiducial cross section results, where the fiducial volume is defined from the single-muon \pt and \etalab selection, are shown in Figs.~\ref{fig:xsec_noacc_mass} and \ref{fig:xsec_noacc_other}, as functions of the dressed lepton kinematic variables (as discussed in Section~\ref{sec:FSR}), together with the expectations from \POWHEG, using the CT14~\cite{ct14} or CT14+EPPS16~\cite{epps16} PDF sets. Cross sections in the full phase space, $-2.87<\ycm<1.93$, \ie including the acceptance correction for the single-muon kinematic selections, are presented in Figs.~\ref{fig:xsec_mass} and \ref{fig:xsec_other}. 

The CT14+EPPS16 predictions suffer from a larger uncertainty than CT14 alone, which is coming
from the parametrisation of the nuclear modification of the PDFs. Since the dimuon rapidity is strongly correlated with the longitudinal momentum fraction $x_\myPb$ of the parton in the lead nucleus, one can identify 
the shadowing region in the rapidity dependence of the cross section, in the full measured rapidity range for $15<\mmumu<60\GeV$ and at positive rapidity for $60<\mmumu<120\GeV$. In the latter mass range, rapidities
$\ycm \lesssim -1$ correspond to the antishadowing region. The inclusion of EPPS16 nuclear PDF modifications tends to provide a better description of the rapidity dependence in data for $60<\mmumu<120\GeV$ than the 
use of the CT14 PDF alone. Uncertainties in the measurement are also smaller than nPDF uncertainties in the \PZ boson mass region for most analysis bins, showing that these data will impose strong constraints if included in future nPDF fits. 

The mass dependence of the cross section sheds further light on the shadowing effects probed at low mass, \ie at lower $x_\myPb$ and lower scales than using \PZ bosons. The cross section
measurement extends down to masses close to the \PgU meson masses, with potential implications in the understanding of the interplay between nPDF and other effects in quarkonium production in 
proton-nucleus collisions~\cite{Arleo:2015qiv}.

The difference between the fiducial cross sections, shown in Figs.~\ref{fig:xsec_noacc_mass} and \ref{fig:xsec_noacc_other}, and the ones corrected to the full phase space, shown in Figs.~\ref{fig:xsec_mass} and \ref{fig:xsec_other}, is largest for low masses. The absence of acceptance correction in the former results reduces their model dependence and corresponding theoretical uncertainty, making clearer the trend for a higher cross section in data for low dimuon masses compared to the \POWHEG expectation.

The \pt and \phistar dependencies of the cross section, especially in the \PZ boson mass region, both point to a slight mismodelling in \POWHEG, reminiscent of the trend reported previously~\cite{hin-15-002}, where the data are softer than \POWHEG predictions. The large sensitivity of these observables to the details of the QCD model, especially nonperturbative effects, is also observed in \pp collisions~\cite{Sirunyan:2019bzr} and prevents one from using them to draw unambiguous conclusions about nPDFs.
This precise measurement in \pPb collisions provides new insight into the soft QCD phenomena dominating the production at low boson \pt or \phistar, and their possible modification with respect to \pp collisions.

Integrated cross sections are also reported, in two mass ranges, in the fiducial region (fid.) or in the full phase space for $-2.87<\ycm<1.93$ (full):
\begin{linenomath}
   \begin{equation}
   \begin{aligned}
      \sigma (\pPb \to \gamma^* / \PZ \to \PGmp\PGmm \textrm{, fid., } 15 < \mmumu < 60\GeV) & =  22.6 \pm 0.5 \stat \pm 0.8 \syst\unit{nb}, \nonumber \\ 
      \sigma (\pPb \to \gamma^* / \PZ \to \PGmp\PGmm  \textrm{, fid., } 60 < \mmumu < 120\GeV) & =  122.3 \pm 0.9 \stat \pm 1.6 \syst\unit{nb}, \nonumber  \\
      \sigma (\pPb \to \gamma^* / \PZ \to \PGmp\PGmm  \textrm{, full, } 15 < \mmumu < 60\GeV) & =  181.7 \pm 3.6 \stat \pm 14.4 \syst\unit{nb}, \nonumber \\ 
      \sigma (\pPb \to \gamma^* / \PZ \to \PGmp\PGmm  \textrm{, full, } 60 < \mmumu < 120\GeV) & =  177.7 \pm 1.3 \stat \pm 2.7 \syst\unit{nb}. \nonumber 
   \end{aligned}
   \end{equation}
\end{linenomath}

In Tables~\ref{tab:chi2_noacc} and \ref{tab:chi2}, the $\chi^2$ values between the data and the predictions are reported, accounting for the bin-to-bin correlations for experimental (systematic uncertainties, shown in Figs.~\ref{fig:cormat_mass} and \ref{fig:cormat_other}) and theoretical (from nPDF) uncertainties. The observations discussed above from Figs.~\ref{fig:xsec_noacc_mass} to \ref{fig:xsec_other} can be made here more quantitatively and more precisely with fiducial cross sections, thanks to the smaller systematic uncertainty. The inclusion of the EPPS16 modifications to the PDFs of the lead nucleus tends to improve the description for \ycm in the \PZ boson mass region, but conclusions are not clear for other quantities, and could even be opposite in the case of \pt and \phistar in that region. However, the manifestly imperfect modelling of the cross sections in \POWHEG prevents from drawing strong conclusions about nPDFs using these variables. 

\begin{table}[htb!]
 \topcaption{
 \label{tab:chi2_noacc}
 $\chi^2$ values between the data and the \POWHEG predictions and associated probability, from the fiducial cross sections, when experimental and theoretical bin-to-bin correlations are taken into account. The integrated luminosity uncertainty is included in the experimental uncertainties.
 }
 \centering
\begin{tabular}{lcccccccc}
   \multirow{2}{*}{Observable} & \multirow{2}{*}{Mass range} & & \multicolumn{3}{c}{CT14} & \multicolumn{3}{c}{EPPS16} \\
   & & & $\chi^{2}$ & dof & Prob. [\%] & $\chi^{2}$ & dof & Prob. [\%] \\
\hline
   \mmumu & $15<\mmumu<600\GeV$ & & 35 & 13 & 0.10 & 30 & 13 & 0.42 \\
   \ycm &$60 < \mmumu < 120\GeV$ & & 51 & 24 & 0.12 & 35 & 24 & 6.6 \\
   \pt &$60 < \mmumu < 120\GeV$ & & 26 & 17 & 8.4 & 52 & 17 & 0.002 \\
   \phistar &$60 < \mmumu < 120\GeV$ & & 23 & 17 & 17 & 45 & 17 & 0.03 \\
   \ycm &$15 < \mmumu < 60\GeV$ & & 11 & 12 & 50 & 10 & 12 & 58 \\
   \pt &$15 < \mmumu < 60\GeV$ & & 12 & 8 & 15 & 8.5 & 8 & 38 \\
   \phistar &$15 < \mmumu < 60\GeV$ & & 8.3 & 9 & 50 & 9.0 & 9 & 44 \\
\end{tabular}
\end{table}

\begin{table}[htb!]
 \topcaption{
 \label{tab:chi2}
 $\chi^2$ values between the data and the \POWHEG predictions and associated probability, from the full phase space cross sections, when experimental and theoretical bin-to-bin correlations are taken into account. The integrated luminosity uncertainty is included in the experimental uncertainties.
 }
 \centering
\begin{tabular}{lcccccccc}
   \multirow{2}{*}{Observable} & \multirow{2}{*}{Mass range} & & \multicolumn{3}{c}{CT14} & \multicolumn{3}{c}{EPPS16} \\
   & & & $\chi^{2}$ & dof & Prob. [\%] & $\chi^{2}$ & dof & Prob. [\%] \\
\hline
   \mmumu & $15<\mmumu<600\GeV$ & & 27 & 13 & 1.2 & 25 & 13 & 2.0 \\
   \ycm &$60 < \mmumu < 120\GeV$ & & 50 & 24 & 0.13 & 35 & 24 & 7.3 \\
   \pt &$60 < \mmumu < 120\GeV$ & & 28 & 17 & 4.5 & 51 & 17 & 0.003 \\
   \phistar &$60 < \mmumu < 120\GeV$ & & 25 & 17 & 9.3 & 44 & 17 & 0.03 \\
   \ycm &$15 < \mmumu < 60\GeV$ & & 7.4 & 12 & 83 & 6.0 & 12 & 92 \\
   \pt &$15 < \mmumu < 60\GeV$ & & 14 & 8 & 8.3 & 8.3 & 8 & 40 \\
   \phistar &$15 < \mmumu < 60\GeV$ & & 6.2 & 9 & 72 & 6.4 & 9 & 69 \\
\end{tabular}
\end{table}

\begin{figure}[htb]
 {\centering
\includegraphics[width=0.49\textwidth]{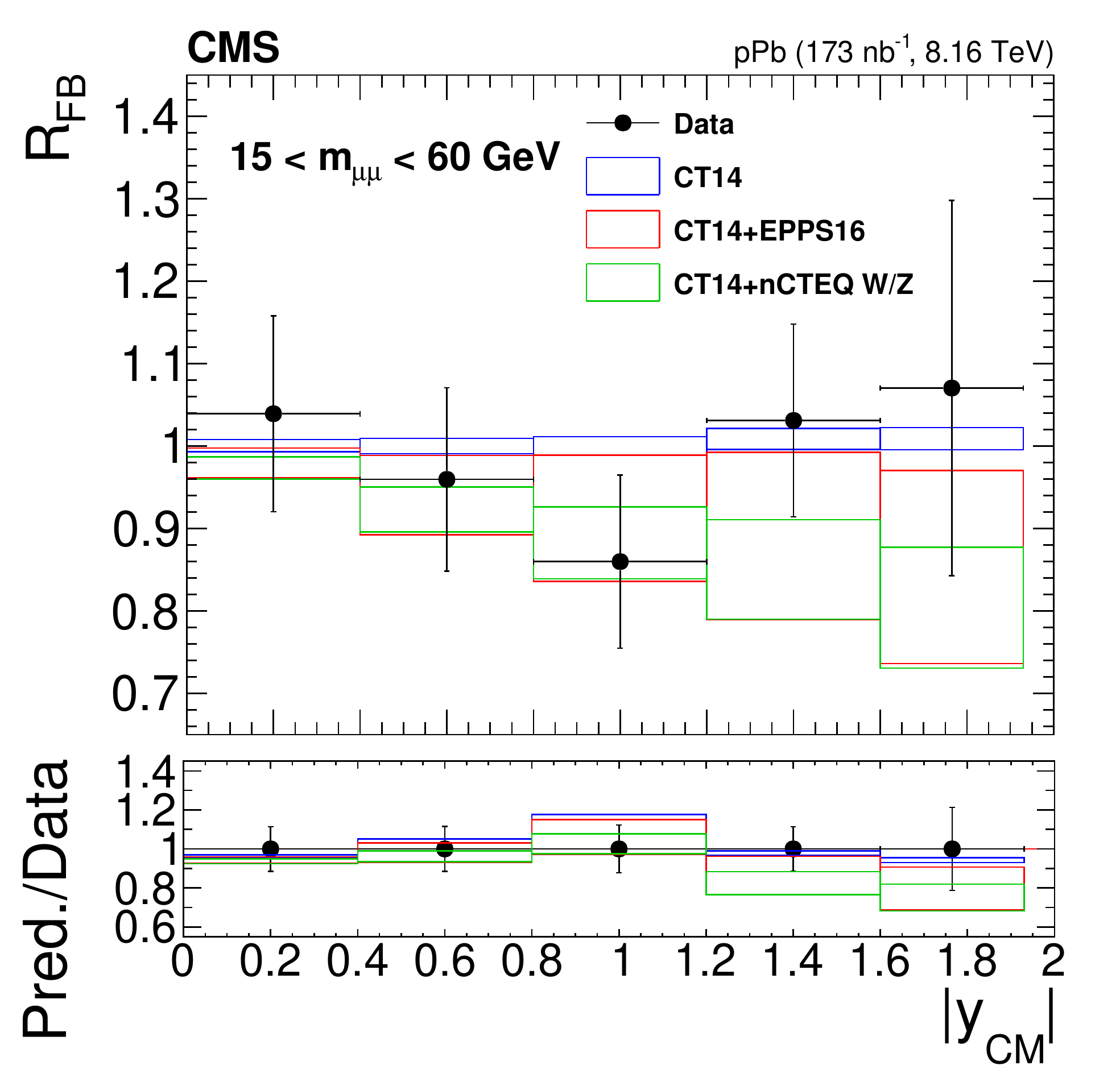} 
\includegraphics[width=0.49\textwidth]{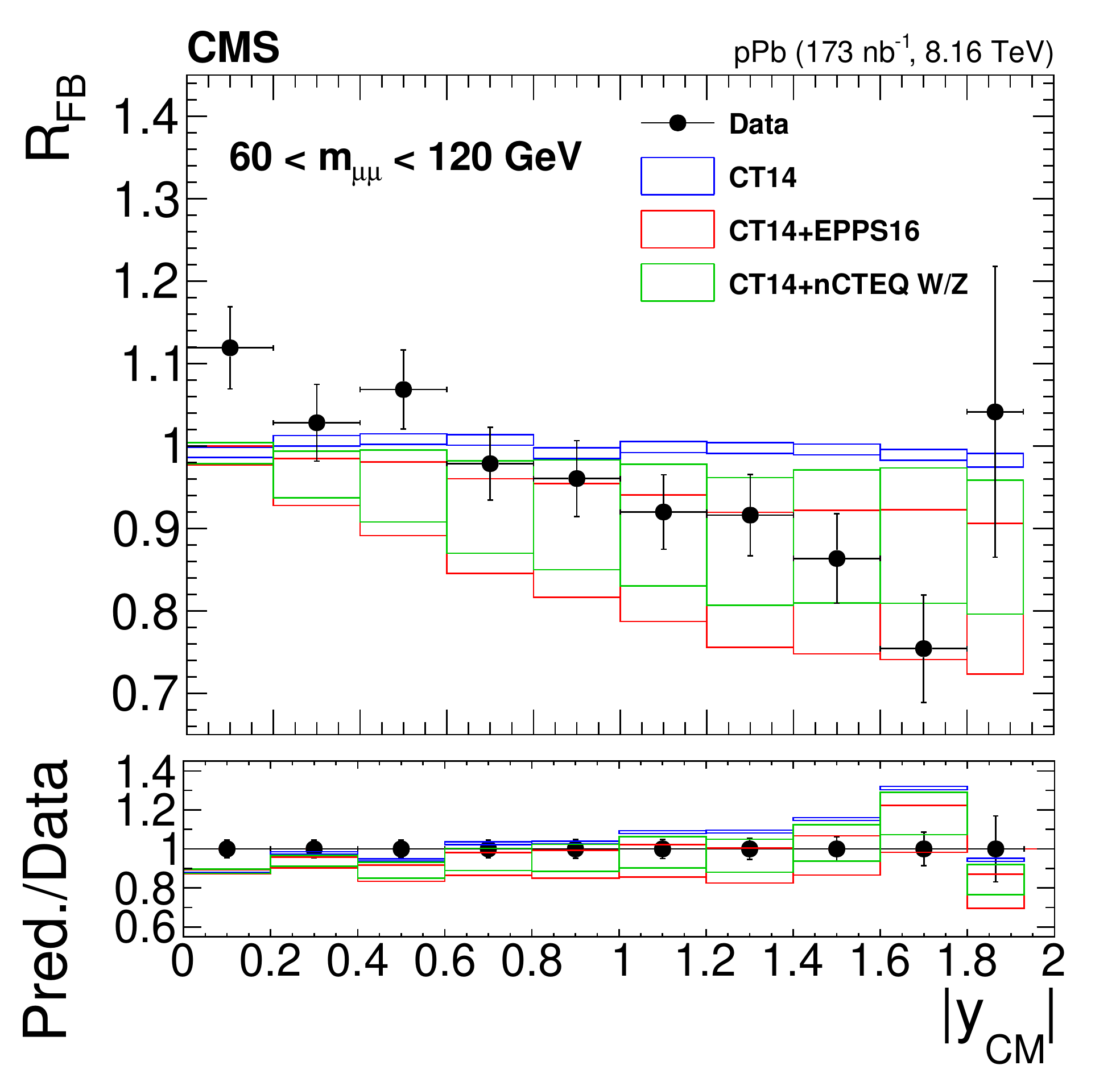}
\caption{
\label{fig:rfb}
Forward-backward ratios
for $15<\mmumu<60\GeV$ (left) and $60<\mmumu<120\GeV$ (right).
The error bars on the data points represent the quadratic sum of the statistical and systematic uncertainties. The theory predictions from the \POWHEG NLO generator
   are also shown, using CT14~\cite{ct14} (blue), CT14+EPPS16~\cite{epps16} (red), or CT14+nCTEQ15WZ~\cite{Kusina:2020lyz} (green) PDF sets. The boxes show the 68\% confidence level (n)PDF uncertainty in these predictions. The ratios of predictions over data are shown in the lower panels, where the data and the (n)PDF uncertainties are shown separately, as error bars around one and as coloured boxes, respectively.
}
}
\end{figure}

Forward-backward ratios ($\rfb$) are built from the rapidity-dependent cross sections in the two mass regions, defined as the ratio of the $\ycm>0$ to the $\ycm<0$ cross sections (\Pp-going to \myPb-going). They are shown in \fig{fig:rfb}. In both mass regions, the $\rfb$ is by construction equal to unity in the absence of nuclear effect (CT14), but decreasing with $\abs{\ycm}$ with CT14+EPPS16 and CT14+nCTEQ15WZ~\cite{Kusina:2020lyz}. Similar conclusions are drawn as from the rapidity
dependence of the cross section, but the construction of these ratios allows for the partial cancellation of theoretical and experimental uncertainties, accounting for the correlations
described in the previous section. In particular, for $60<\mmumu<120\GeV$ and at large $\abs{\ycm}$, an indication of a forward-backward ratio smaller than unity is found, consistent
with the expectation from the combination of shadowing and antishadowing effects expected with CT14+EPPS16, as well as with similar results from \PW bosons~\cite{Sirunyan:2019dox}. Predictions using CT14+nCTEQ15WZ are found to be in good agreement with the data. The larger amount of shadowing in nCTEQ15~\cite{Kovarik:2015cma}, hinted by the recent \PW boson measurement~\cite{Sirunyan:2019dox}, is not predicted with nCTEQ15WZ. The low mass region is less conclusive, but nPDF uncertainties are smaller in this selection for nCTEQ15WZ than for EPPS16. Finally, experimental uncertainties for $60<\mmumu<120\GeV$ are smaller than the nPDF ones, once again showing relevance of these data to the study of nPDF effects.

\section{Summary}
\label{sec:summary}

Differential cross section measurements of the Drell--Yan process in the dimuon channel in proton-lead collisions at $\sqrtsNN = 8.16\TeV$ have been reported, including the transverse momentum (\pt) and rapidity dependencies in the \PZ boson mass
region ($60<\mmumu<120\GeV$).
In addition, for the first time in collisions including nuclei, the \pt and rapidity dependence for smaller masses
$15<\mmumu<60\GeV$ have been measured. The dependence with \phistar (a geometrical variable that highly correlates with dimuon \pt but is determined with higher precision) for both $15<\mmumu<60\GeV$ and $60<\mmumu<120\GeV$ and the mass dependence from 15 to 600\GeV have been presented, also for the first time in proton-nucleus collisions. 
Finally, forward-backward ratios have been built from the rapidity-dependent cross sections for $\ycm>0$ to $\ycm<0$ in both mass regions, highlighting the presence of nuclear effects in the parton distribution functions. 

Results for $60<\mmumu<120\GeV$ are the most precise to date, featuring smaller uncertainties than the theoretical predictions, and provide novel constraints on the quark and antiquark nuclear parton distribution functions (nPDFs). 
Measurements in the lower mass range $15<\mmumu<60\GeV$ give access to a new phase space for nPDF studies, extending to lower longitudinal momentum fraction $x$ and lower energy scale $Q^2$.
The \pt- and \phistar-dependent results are also very sensitive to the details of model details, such as soft quantum chromodynamics phenomena, which they may help to better understand in \pPb collisions.

\begin{acknowledgments}
   We congratulate our colleagues in the CERN accelerator departments for the excellent performance of the LHC and thank the technical and administrative staffs at CERN and at other CMS institutes for their contributions to the success of the CMS effort. In addition, we gratefully acknowledge the computing centres and personnel of the Worldwide LHC Computing Grid and other centres for delivering so effectively the computing infrastructure essential to our analyses. Finally, we acknowledge the enduring support for the construction and operation of the LHC, the CMS detector, and the supporting computing infrastructure provided by the following funding agencies: BMBWF and FWF (Austria); FNRS and FWO (Belgium); CNPq, CAPES, FAPERJ, FAPERGS, and FAPESP (Brazil); MES (Bulgaria); CERN; CAS, MoST, and NSFC (China); COLCIENCIAS (Colombia); MSES and CSF (Croatia); RIF (Cyprus); SENESCYT (Ecuador); MoER, ERC PUT and ERDF (Estonia); Academy of Finland, MEC, and HIP (Finland); CEA and CNRS/IN2P3 (France); BMBF, DFG, and HGF (Germany); GSRT (Greece); NKFIA (Hungary); DAE and DST (India); IPM (Iran); SFI (Ireland); INFN (Italy); MSIP and NRF (Republic of Korea); MES (Latvia); LAS (Lithuania); MOE and UM (Malaysia); BUAP, CINVESTAV, CONACYT, LNS, SEP, and UASLP-FAI (Mexico); MOS (Montenegro); MBIE (New Zealand); PAEC (Pakistan); MSHE and NSC (Poland); FCT (Portugal); JINR (Dubna); MON, RosAtom, RAS, RFBR, and NRC KI (Russia); MESTD (Serbia); SEIDI, CPAN, PCTI, and FEDER (Spain); MOSTR (Sri Lanka); Swiss Funding Agencies (Switzerland); MST (Taipei); ThEPCenter, IPST, STAR, and NSTDA (Thailand); TUBITAK and TAEK (Turkey); NASU (Ukraine); STFC (United Kingdom); DOE and NSF (USA).
    
   \hyphenation{Rachada-pisek} Individuals have received support from the Marie-Curie programme and the European Research Council and Horizon 2020 Grant, contract Nos.\ 675440, 724704, 752730, and 765710 (European Union); the Leventis Foundation; the Alfred P.\ Sloan Foundation; the Alexander von Humboldt Foundation; the Belgian Federal Science Policy Office; the Fonds pour la Formation \`a la Recherche dans l'Industrie et dans l'Agriculture (FRIA-Belgium); the Agentschap voor Innovatie door Wetenschap en Technologie (IWT-Belgium); the F.R.S.-FNRS and FWO (Belgium) under the ``Excellence of Science -- EOS" -- be.h project n.\ 30820817; the Beijing Municipal Science \& Technology Commission, No. Z191100007219010; the Ministry of Education, Youth and Sports (MEYS) of the Czech Republic; the Deutsche Forschungsgemeinschaft (DFG), under Germany's Excellence Strategy -- EXC 2121 ``Quantum Universe" -- 390833306, and under project number 400140256 - GRK2497; the Lend\"ulet (``Momentum") Programme and the J\'anos Bolyai Research Scholarship of the Hungarian Academy of Sciences, the New National Excellence Program \'UNKP, the NKFIA research grants 123842, 123959, 124845, 124850, 125105, 128713, 128786, and 129058 (Hungary); the Council of Science and Industrial Research, India; the Ministry of Science and Higher Education and the National Science Center, contracts Opus 2014/15/B/ST2/03998 and 2015/19/B/ST2/02861 (Poland); the National Priorities Research Program by Qatar National Research Fund; the Ministry of Science and Higher Education, project no. 0723-2020-0041 (Russia); the Programa Estatal de Fomento de la Investigaci{\'o}n Cient{\'i}fica y T{\'e}cnica de Excelencia Mar\'{\i}a de Maeztu, grant MDM-2015-0509 and the Programa Severo Ochoa del Principado de Asturias; the Thalis and Aristeia programmes cofinanced by EU-ESF and the Greek NSRF; the Rachadapisek Sompot Fund for Postdoctoral Fellowship, Chulalongkorn University and the Chulalongkorn Academic into Its 2nd Century Project Advancement Project (Thailand); the Kavli Foundation; the Nvidia Corporation; the SuperMicro Corporation; the Welch Foundation, contract C-1845; and the Weston Havens Foundation (USA).
\end{acknowledgments}

\bibliography{auto_generated} 

\cleardoublepage \appendix\section{The CMS Collaboration \label{app:collab}}\begin{sloppypar}\hyphenpenalty=5000\widowpenalty=500\clubpenalty=5000\vskip\cmsinstskip
\textbf{Yerevan Physics Institute, Yerevan, Armenia}\\*[0pt]
A.M.~Sirunyan$^{\textrm{\dag}}$, A.~Tumasyan
\vskip\cmsinstskip
\textbf{Institut f\"{u}r Hochenergiephysik, Wien, Austria}\\*[0pt]
W.~Adam, F.~Ambrogi, T.~Bergauer, M.~Dragicevic, J.~Er\"{o}, A.~Escalante~Del~Valle, R.~Fr\"{u}hwirth\cmsAuthorMark{1}, M.~Jeitler\cmsAuthorMark{1}, N.~Krammer, L.~Lechner, D.~Liko, T.~Madlener, I.~Mikulec, F.M.~Pitters, N.~Rad, J.~Schieck\cmsAuthorMark{1}, R.~Sch\"{o}fbeck, M.~Spanring, S.~Templ, W.~Waltenberger, C.-E.~Wulz\cmsAuthorMark{1}, M.~Zarucki
% ...
\vskip\cmsinstskip
\dag: Deceased\\
1:  Also at Vienna University of Technology, Vienna, Austria\\
% ...
\end{sloppypar}
\end{document}